\documentclass[aps,prb,reprint,superscriptaddress]{revtex4-1}
\usepackage{graphicx}
\usepackage[usenames,dvipsnames]{color}
\usepackage[caption=false]{subfig}
\graphicspath{ {Figures/} }

\begin{document}


\title{Crystallographic investigation of Au nanoparticles embedded in a SrTiO$_3$ thin film for plasmonics applications by means of synchrotron radiation}


\author{Davide Pincini}

\affiliation{European Synchrotron Radiation Facility, CS 40220, 71, avenue des Martyrs, F-38043 Grenoble Cedex 9, France
}
\affiliation{Dipartimento di Fisica, Politecnico di Milano, Piazza Leonardo Da Vinci 32, 20133 Milano, Italy
}

\author{Claudio Mazzoli}
\affiliation{Dipartimento di Fisica, Politecnico di Milano, Piazza Leonardo Da Vinci 32, 20133 Milano, Italy
}

\author{Hendrik Bernhardt}
\affiliation{Institut f$\ddot{\textrm{u}}$r Festk$\ddot{\textrm{o}}$rperphysik, Friedrich-Schiller-Universit$\ddot{\textrm{a}}$t Jena, Helmholtzweg 5, 07743 Jena, Germany
}

\author{Christian Katzer}
\affiliation{Institut f$\ddot{\textrm{u}}$r Festk$\ddot{\textrm{o}}$rperphysik, Friedrich-Schiller-Universit$\ddot{\textrm{a}}$t Jena, Helmholtzweg 5, 07743 Jena, Germany
}

\author{Frank Schmidl}
\affiliation{Institut f$\ddot{\textrm{u}}$r Festk$\ddot{\textrm{o}}$rperphysik, Friedrich-Schiller-Universit$\ddot{\textrm{a}}$t Jena, Helmholtzweg 5, 07743 Jena, Germany
}

\author{Ingo Uschmann}
\affiliation{Institut f$\ddot{\textrm{u}}$r  Optik und Quantenelektronik, Friedrich-Schiller-Universit$\ddot{\textrm{a}}$t Jena, Max-Wien-Platz 1, 07743 Jena, Germany
}

\author{Carsten Detlefs}
\affiliation{European Synchrotron Radiation Facility, CS 40220, 71, avenue des Martyrs, F-38043 Grenoble Cedex 9, France
}


\date{\today}

\begin{abstract}
Self-organized monocrystalline Au nanoparticles with potential applications in plasmonics are grown in a SrTiO$_3$ matrix by a novel two-step deposition process. The crystalline preferred orientation of these Au nanoparticles is investigated by synchrotron hard x-ray diffraction. Nanoparticles preferentially align with the (111) direction along the substrate normal (001), whereas two in-plane orientations are found with $[110]_{SrTiO_3} \: || \: [110]_{Au}$ and $[100]_{SrTiO_3} \: || \: [110]_{Au}$. Additionally, a smaller diffraction signal from nanoparticles with the (001) direction parallel to the substrate normal (001) is observed; once again, two in-plane orientations are found, with $[100]_{SrTiO_3} \: || \: [100]_{Au}$ and $[100]_{SrTiO_3} \: || \: [110]_{Au}$. The populations of the two in-plane orientations are found to depend on the thickness of the gold film deposited in the first step of the growth. 

\end{abstract}


\maketitle

\section{Introduction}
\begin{figure*}
	\includegraphics[trim = 0mm 0mm 0mm 0mm,width=0.7\textwidth]{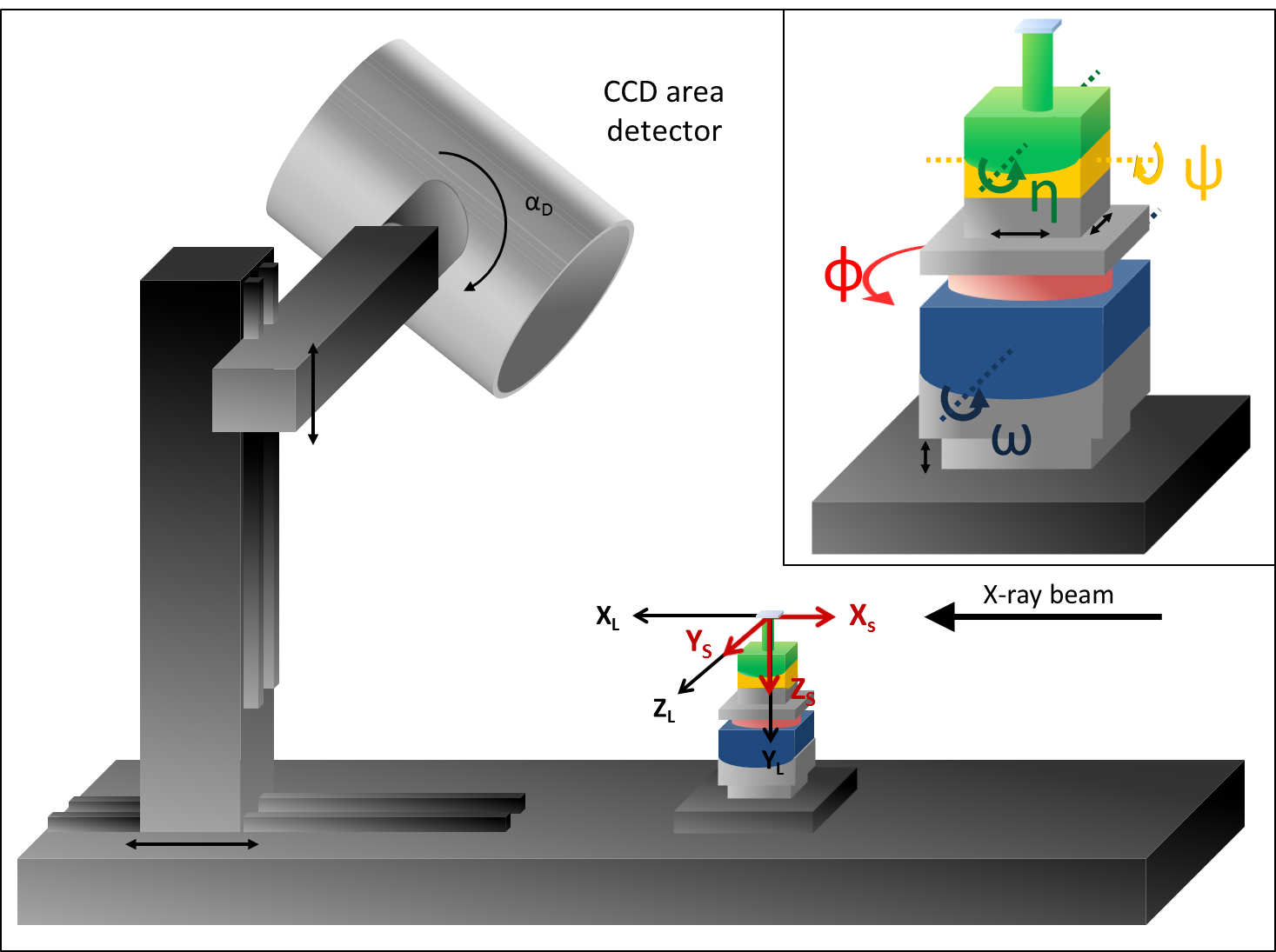}
	\caption{\label{CCD_setup}(Color online) Schematic of the CCD area detector setup in the experimental hutch n$^\circ$1 of beamline ID06 of the European Synchrotron Radiation Facility (ESRF). The laboratory coordinate system $X_LY_LZ_L$ and the sample coordinate system $X_SY_SZ_S$ (with all the sample rotations set to zero) as defined by \citet{he2011two} are displayed. In the top right box, the 4 rotations of the sample mounting are shown in their zero position.}	
\end{figure*}
The optical properties of metallic nanostructures are governed by the localized surface plasmon resonance (LSPR) \cite{kelly2003optical,link2003optical,link1999spectral,vollmer1995optical}. The LSPR can be tuned via the morphological and structural features of the nanoparticles and thus presents considerable potential for scientific and technological applications. The sensitivity of the plasmon resonance and the associated local field enhancement on the medium surrounding the nanoparticles makes them suitable e.g.\ as sensors for the detection of biological analytes \cite{prasad2004nanophotonics}, Surface Enhanced Raman Spectroscopy (SERS) \cite{chang1982surface}, radiative decay engineering \cite{lakowicz2001radiative} of nearby fluorescent molecules, sub-wavelength optical wave guides \cite{maier2003local}, apertureless near-field microscopy and nanostructure photofabrication \cite{prasad2004nanophotonics}. Among the applications not directly related to the LSPR, nanometer-sized Au particles have been proven to act as highly efficient catalysts \cite{chen2004structure,bell2003impact,lopez2004origin,valden1998onset}. 

To date, most methods for the fabrication of metallic nanoparticles rely on lithographic techniques or various chemical processes \cite{pelton2008metal}. Lithographic techniques allow a large number of 2D structures with a variety of different shapes to be produced. Nevertheless the nanostructures are generally polycrystalline with not well controlled grain size, orientation and arrangement: this leads to a great surface roughness and size and shape dispersion. Different chemical methods have been used as an alternative to lithography. Although they guarantee the production of single crystal nanoparticles with nearly atomically smooth surfaces, they are in this case difficult to arrange in a predefined pattern and are normally randomly distributed. The samples investigated in this study were grown using a novel two-step process \cite{christke2011optical,katzer2013matrix} which consists of a deposition of a Au layer (the so called seed layer) on a polished single crystal 5 x 10 x 1 mm$^3$ SrTiO$_3$(001) substrate in an Ar atmosphere of $16\cdot10^{-4}$ Torr followed by an annealing at 1070K for 5 minutes and a subsequent deposition of a SrTiO$_3$ thin film in an oxygen atmosphere of 100 Pa. During the latter deposition step, the Au layer (dewetted after the annealing process to form 3D Au islands) self-assembles to homogeneously arranged monocrystalline Au nanoparticles of anisotropic shape, mostly located at a characteristic depth within the STO thin film \cite{christke2011optical}. The size and shape of the nanoparticles
can be tuned by varying the deposited amount of Au and STO: their optical properties can thus be readily tuned \cite{christke2011optical,katzer2013matrix}. Furthermore, the possibility of a selective etching of the STO matrix has already been proved \cite{katzer2013matrix}, thus allowing to expose the nanoparticles to air. Therefore, this growth process represents a valid alternative to other more traditional fabrication methods, particularly for the production of plasmonics active sensors in life sciences.

Despite the potential just pointed out, little is known about the growth process leading to the Au nanoparticles formation as well as the structural properties of both the nanoparticles and the STO thin film. These properties are not only important for the potential practical applications of this kind of systems, but are also of great relevance to gain a deeper fundamental understanding of the physics involved in the growth of metallic nanostructures (and, more generally, of metallic thin films) on ceramic substrates, which is still not widely explored \cite{francis2007crystal,fu2007interaction,silly2006bimodal} (especially for the case of the Au). 
The present work aims to fill this gap of knowledge by investigating the crystallographic properties of such a system by means of x-ray diffraction, in particular the preferred crystallographic orientation of the embedded Au nanoparticles. For this purpose, samples with a Au layer thickness ranging in the interval 1-8 nm and a fixed STO thin film thickness of 275 nm were probed. A sample grown in the same way (5 nm of Au), but without the STO top layer was additionally considered: it allowed to investigate the 3D islands of the Au layer dewetted by the annealing process. The STO thin film was found to display a crystalline growth ((001) epitaxy) for all the values of the Au layer thickness below 8 nm: for 8 nm of deposited Au a rather polycristalline film results instead \cite{rawdata}. 

\section{Experimental setup}
The x-ray diffraction measurements were carried out at the beamline ID06 of the European Synchrotron Radiation Facility (ESRF), Grenoble (FR). A first set of measurements (performed in the hutch n$^\circ$1) made use of a focused beam at an energy of 7 keV and a CCD detector mounted in reflection (Bragg) geometry (Fig.~\ref{CCD_setup}). 
The relatively low photon energy and the reflection geometry were chosen to limit the penetration into the substrate, and to thus optimize the ratio of signal from the near-surface nanoparticles vs.\ background from the substrate.
The sample was mounted on a 4-circle goniometer (see Fig.~\ref{CCD_setup}) and diffraction data were recorded for multiple sample positions to investigate the texture of the Au nanoparticles. The data acquisition relied on two main types of goniometer scans. In the first one (used to obtain the pole figures discussed in Section~\ref{results_discussion}) different values of the $\omega$ angle were scanned over a wide range, keeping the $\phi$ axis fixed. In the other one (used to study the in-plane orientation discussed in Section~\ref{in_plane}) different values of the $\phi$ angle were scanned over a wide range, keeping the $\omega$ axis fixed. The diffraction spots on the 2D detector were indexed using the mathematical framework described by \citet{he2011two}, adapted to the present geometry. For a spot of diffracted light in a position $(x,y)$ on the detector surface,  the corresponding unit diffraction vector $\mathbf{h}_L=(\mathbf{K}_{out}-\mathbf{K}_{in})/(|\mathbf{K}_{out}-\mathbf{K}_{in}|)$ in the laboratory coordinate system $X_LY_LZ_L$ can be calculated once known the detector position. For each position of the sample the corresponding vector $\mathbf{h}_S$ in the sample coordinate system $X_SY_SZ_S$ is given by
\begin{equation}
\label{eq:lab_to_sample_mod}
\mathbf{h}_S=\mathbf{B}\:\mathbf{U}\:\mathbf{H}(\eta)\:\mathbf{\Psi}(\psi)\:\mathbf{\Phi}(\phi)\:\mathbf{\Omega}(\omega)\:\mathcal{A}\:\mathbf{h}_L
\end{equation}
where $\mathbf{B}$ depends on the crystal symmetry ($\mathbf{B}=\mathbf{I}$ for the case of a cubic crystal considered), $\mathbf{U}$ takes into account the offsets in the sample mounting and $\mathbf{H}$, $\mathbf{\Psi}$, $\mathbf{\Phi}$ and $\mathbf{\Omega}$ are the rotational matrices associated to the four sample rotational degrees of freedom. The matrix $\mathcal{A}$ corresponds to the fact that, with all the previous matrices equal to the identity $\mathbf{I}$, the sample and the laboratory coordinate system do not coincide within the conventions chosen. 

Despite providing a general overview of the nanoparticles texture properties, the CCD setup 4-circle goniometer does not permit an exact determination of the offsets in the sample mounting. Therefore the absolute orientation of the Au crystals with respect to the substrate cannot be achieved. A second set of measurements (performed in the hutch n$^\circ$2, downstream with respect to n$^\circ$1) was then realized using a single crystal diffractometer in normal beam geometry with a point detector. In this case the offsets in the sample mounting are automatically calculated starting from the position of two known substrate diffraction peaks. The greater distance from the x-ray source imposed the usage of a higher energy (11  keV) beam: given the small field of view of the point detector, the presence of strong substrate peaks due to the higher penetration depth inside the sample did not represent a significant limitation.

\begin{figure}
	\includegraphics[width=0.49\textwidth]{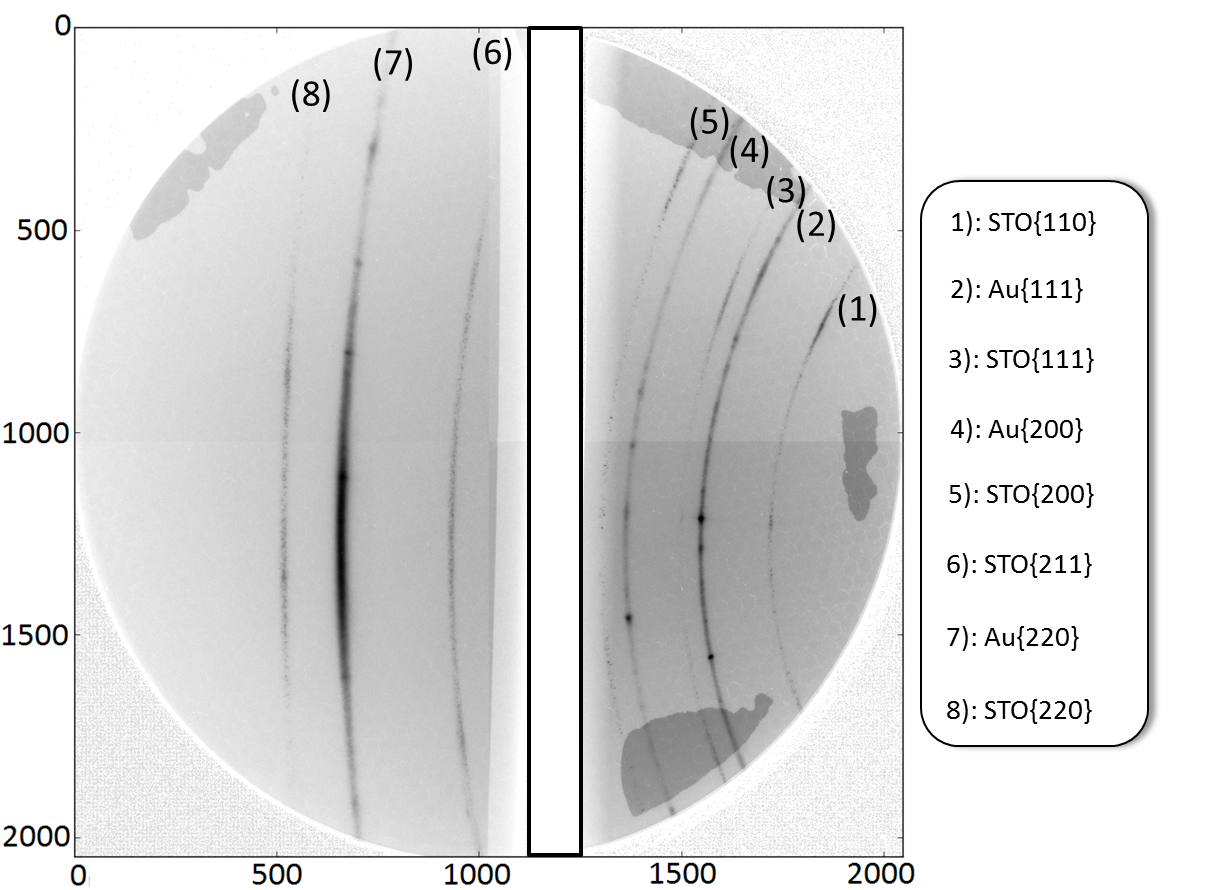}
	\caption{\label{hb19a_3335_NoDark} CCD detector 2048 x 2048 raw image. The diffraction rings exhibit evident texture properties (intensity variation along the rings). The strong STO \{210\} substrate reflection was masked (black rectangle)  to avoid saturation of the detector.}	
\end{figure}
\section{Results and discussion}\label{results_discussion}
The main feature emerging from the diffraction images collected with the CCD area detector setup of Fig.~\ref{CCD_setup} is the presence of a strong texture for the Au nanoparticles. The diffraction rings exhibit bright spots alternated to regions of lower intensity (see Fig.~\ref{hb19a_3335_NoDark}), a clear indication that the Au nanoparticles align along certain crystalline preferred orientations. The latter can be investigated transforming the 2D images of the CCD detector into pole figures, each of which shows the orientation of a selected set of lattice planes $(hkl)$ relative to the substrate. A region of the detector corresponding to the fixed scattering angle $2\theta=\arcsin{\lambda/2d_{hkl}}$ (a diffraction ring) is selected in a series of images with different goniometer settings. Using eq.~(\ref{eq:lab_to_sample_mod}) is then possible to relate the intensity $I$ recorded by every pixel to  the corresponding unit diffraction vector $\mathbf{h}_S$ in the sample coordinate system. The intensity function $I(\mathbf{h}_S)$ can finally be represented on a polar diagram (i.e. the pole figure) by means of a stereographic projection (see \citet{he2011two} for further details).
\begin{figure*}  
	\subfloat[8 nm\label{PoleFigureScan69Au_111_}]{
		\includegraphics[trim=0mm 10mm 0mm -10mm,width=0.49\textwidth]{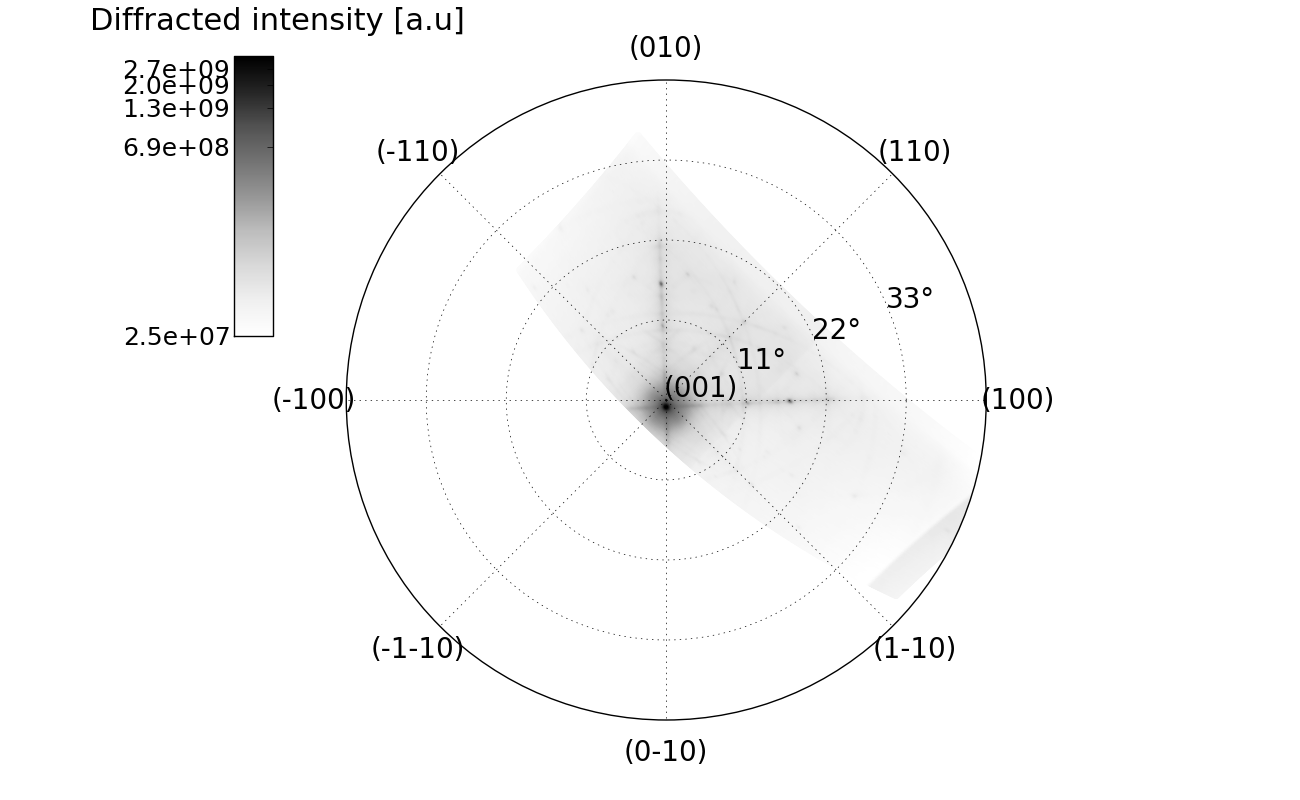}
	}
	\subfloat[5.3 nm\label{PoleFigureScan114Au_111_}]{
		\includegraphics[trim=0mm 10mm 0mm -10mm,width=0.49\textwidth]{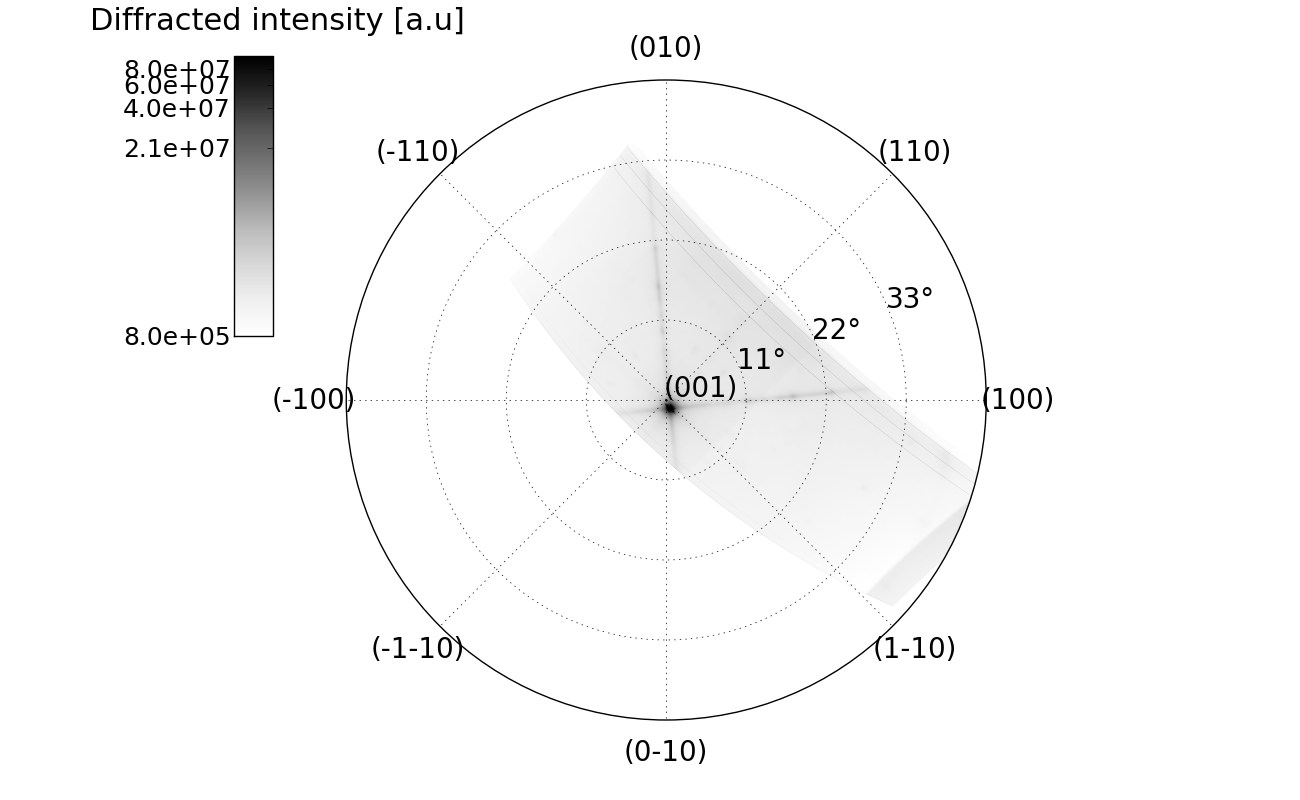}
	}
	\hfill
	\subfloat[4 nm\label{PoleFigureScan39Au_111_}]{
		\includegraphics[trim=0mm 10mm 0mm -10mm,width=0.49\textwidth]{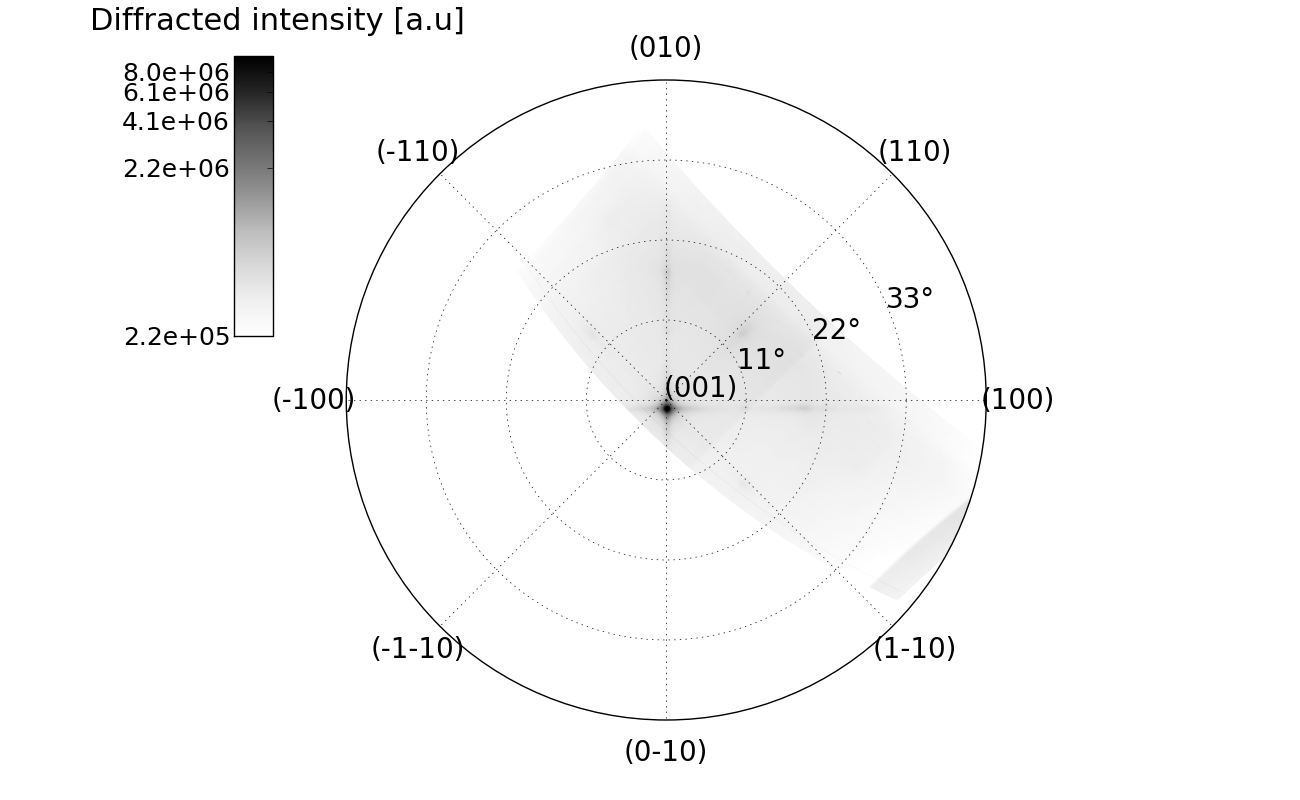}
	}
	\subfloat[3 nm\label{PoleFigureScan169Au_111_}]{
		\includegraphics[trim=0mm 10mm 0mm -10mm,width=0.49\textwidth]{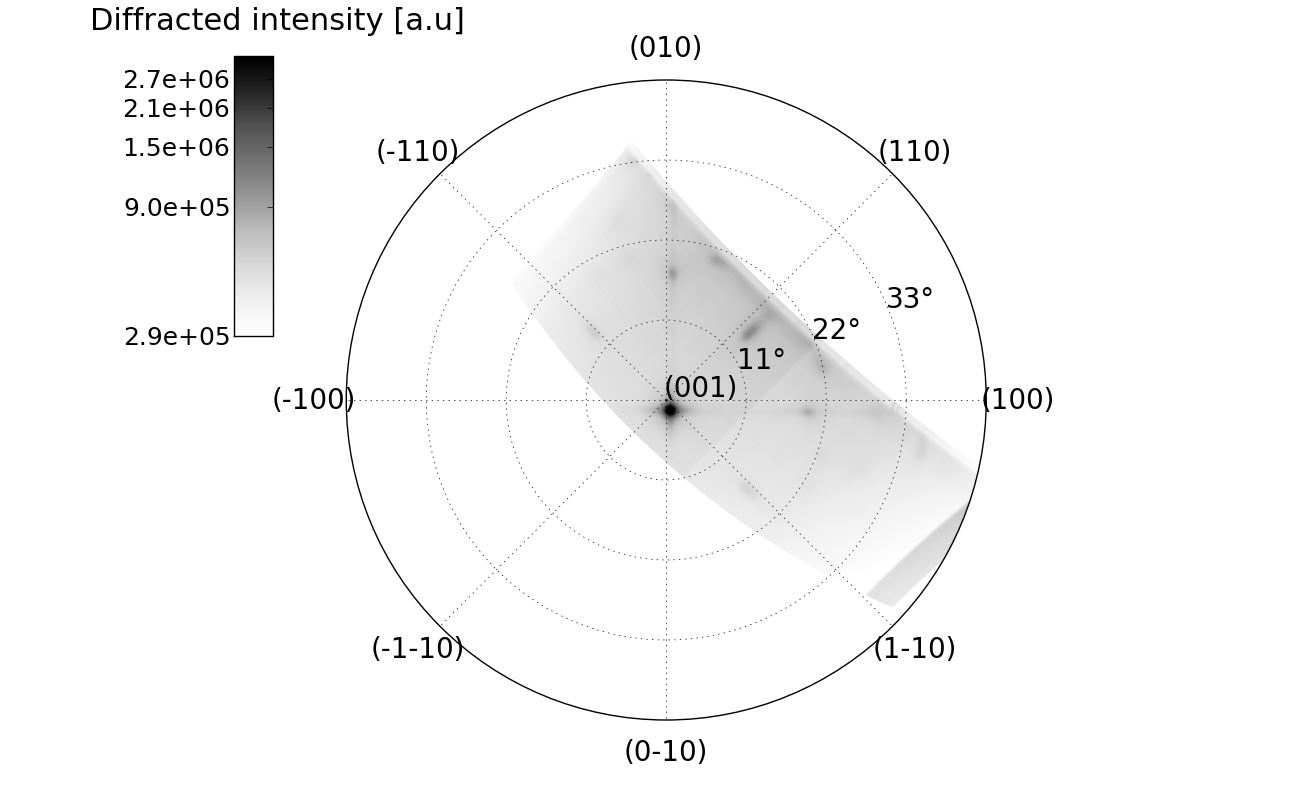}
	}
	\hfill
	\subfloat[2 nm\label{PoleFigureScan141Au_111_}]{
		\includegraphics[trim=0mm 10mm 0mm -10mm,width=0.49\textwidth]{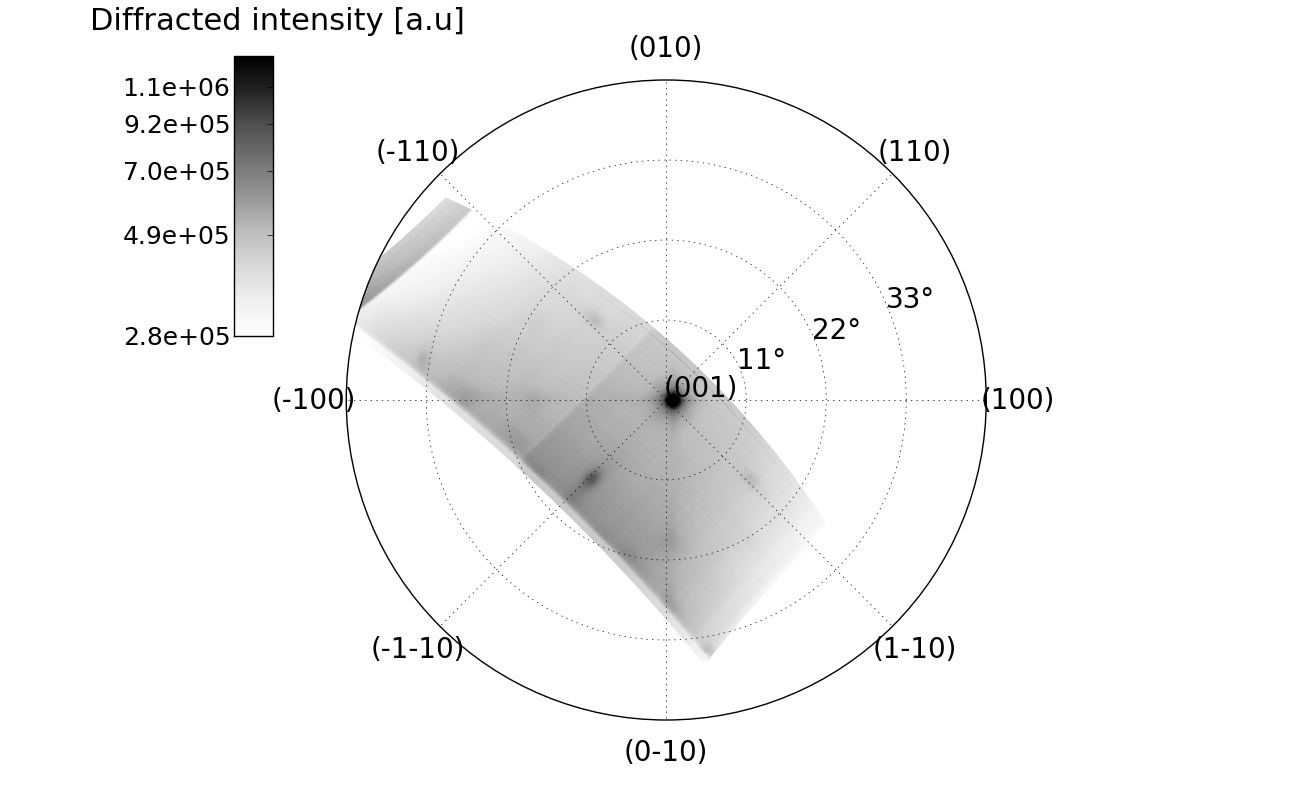}
	}
	\subfloat[1 nm\label{PoleFigureScan99Au_111__cross}]{
		\includegraphics[trim=0mm 10mm 0mm -10mm,width=0.49\textwidth]{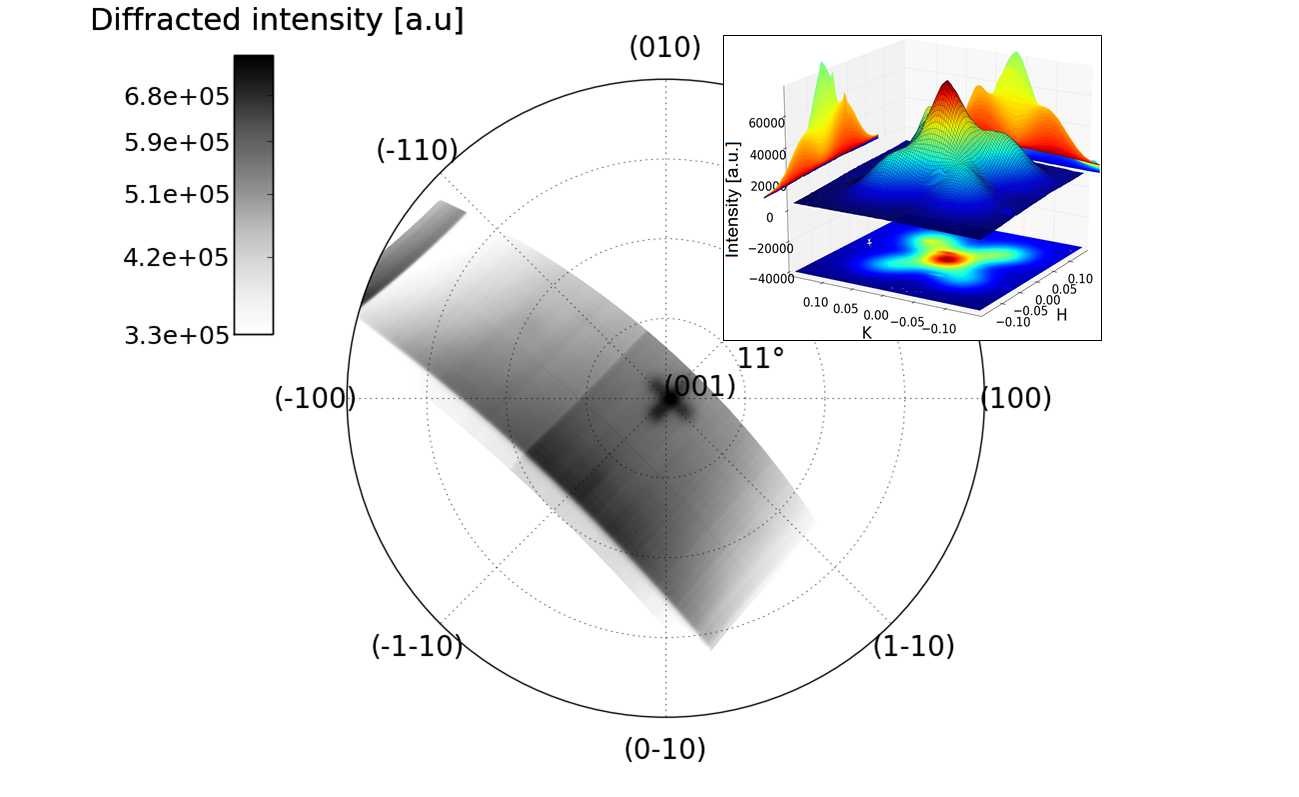}
	}
	\caption{\label{111_pole_figures}(Color online) Au\{111\} pole figures for different values of the Au seed layer thickness. A strong central round-shaped peak is present for the thickness range 2-8 nm. For the 1 nm sample, a cross-like feature appears (see top-right box).}
\end{figure*}
The result is shown in Fig.~\ref{111_pole_figures} for different thicknesses of the Au seed layer in the case of the Au\{111\} reflection. The radial direction is labeled with the values of the angle (referred to as $\chi$) measured with respect to the vertical direction, i.e. the normal to the substrate surface (here indexed as the (001)); for the angular direction the crystallographic directions of the cubic STO substrate have been used ((100), (110), (010), \ldots) rather than the values of the angle measured counterclockwise from the STO(100) (0$^\circ$, 45$^\circ$, 90$^\circ$, \ldots), referred to as $\beta$\cite{he2011two} (which will be used in Section~\ref{in_plane}) . The small angular range of the pole figures probed is simply due to the limited excursion of the motors controlling the sample rotational stages: given the cubic symmetry of the samples, some important conclusions can nevertheless be attained.

\subsection{Orientation along the substrate normal}
The most important aspect shown by Fig.~\ref{111_pole_figures} is the presence of a strong Au\{111\} peak along the direction normal to the substrate surface, i.e. STO(001), which was observed for all the samples probed. It represents a clear proof that a very large number of Au nanoparticles are oriented with the \{111\} crystal direction aligned vertically, as previously observed by \citet{christke2011optical}. An exception is represented by the sample with 1 nm of Au seed layer thickness. Rather than a round-shaped central peak, in this case a cross-like feature, with arms extending along the \{110\} in-plane substrate directions, appears. The presence of this peculiarity was subsequently confirmed by the measurements realized with the single crystal diffractometer (top-right box of Fig.~\ref{PoleFigureScan99Au_111__cross}) and observed for the sample with 1.5 nm of Au as well.
Another aspect shown by the Au\{111\} pole figures is that the central peak for the sample with 8 nm of Au is much more smeared out than for the other samples. 
In general, the situation looks somehow more complex, as the presence of additional lines and small spots (along the (100) and (010) at $\chi\approx11^\circ$ and $\chi\approx16^\circ$) testifies. The observed lack of a rigid crystallographic orientation may be attributed to the greater Au nanoparticles size or the structural modifications of the surrounding STO matrix observed for the relatively high value of the initial Au layer thickness considered \cite{rawdata}. Other Au nanoparticles preferred orientations in addition to the central one are also clearly visible for the samples with 3 and 2 nm of Au. In both cases they cause the presence of spots in the region between $\chi=11^\circ$ and $\chi=22^\circ$ along the (110) and (100) substrate directions. These additional features are less visible in the 4 nm sample and seem not to be present for the samples with 5.3 and 8 nm of Au. For these latter samples, dark continuous lines of diffracted intensity instead appear along directions almost parallel to the (100) and (010) (the small deviation observed could be simply due to the limited precision of the data analysis).

\begin{figure}
	\includegraphics[trim = 0mm 0mm 0mm 0mm,width=0.49\textwidth]{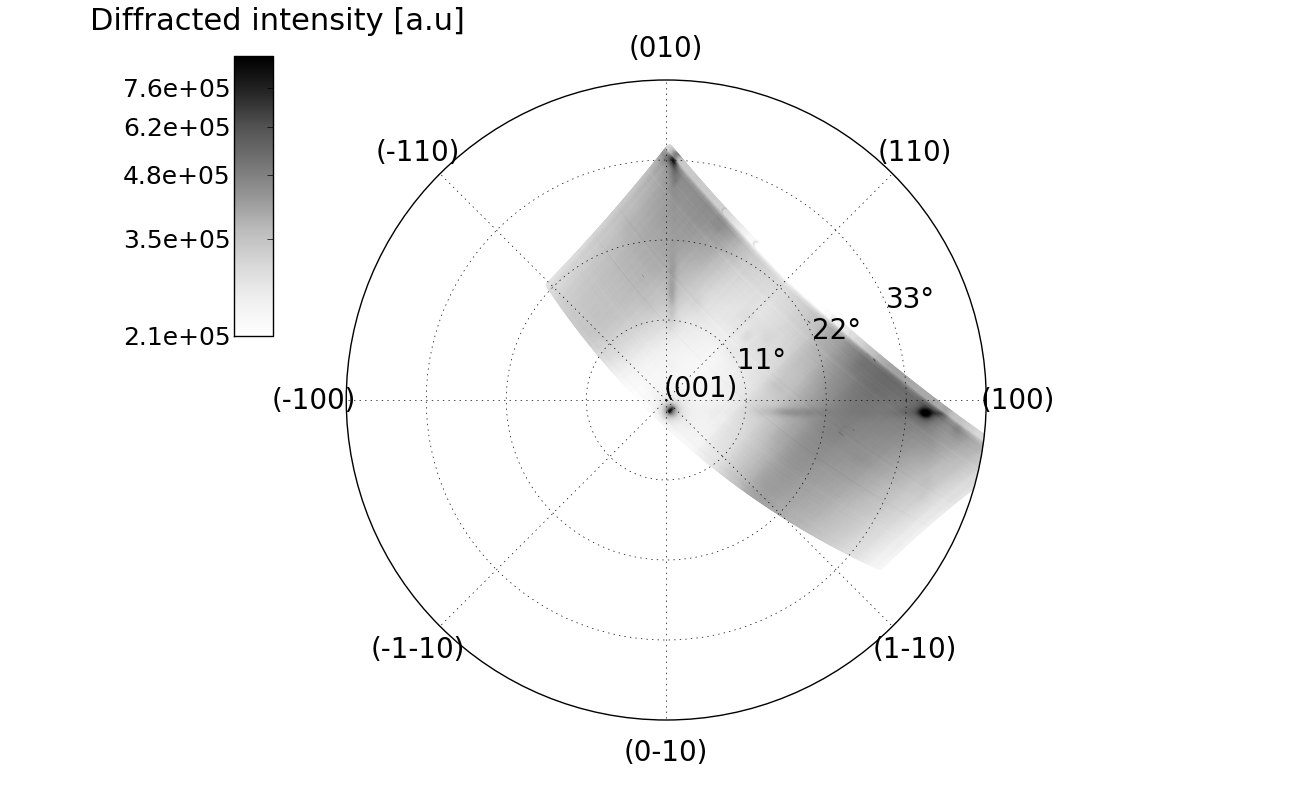}
	\caption{\label{PoleFigureScan169Au_200_}Au\{200\} pole figure for 3 nm of Au seed layer thickness. A central tiny peak as well as additional features along the (100) and (010) directions are present.}	
\end{figure}

In addition to the \{111\}-oriented nanoparticles, there seem to be present a certain fraction of nanoparticles oriented with the \{200\} direction normal to the substrate. All the Au\{200\} pole figures indeed present a tiny peak in the center. As an example the pole figure for the sample with 3 nm of Au seed layer thickness is reported in Fig.~\ref{PoleFigureScan169Au_200_}. As for the Au\{111\} pole figures, other features are present in addition to the central peak. In particular, bright spots are present approximately along the (100) and the (010) substrate directions at a $\chi$ angle around $33^\circ$ for the samples with 8, 5.3 and 3 nm of Au. 
The reader is invited to notice that the out of vertical additional features neither of the Au\{200\} nor of the Au\{111\} pole figures can be straightforwardly attributed to the presence of the other (\{111\} and \{200\}, respectively) Au nanoparticles preferred vertical orientation. This can be easily verified considering the $\chi$ value of the additional spots mentioned above and the angle between the (111) and the (002) lattice directions, equal to $54.74^\circ$.

\begin{figure}  
	\subfloat[Au(111)\label{Au_111__S010}]{
		\includegraphics[trim=0mm 5mm 0mm 0mm,width=0.2\textwidth]{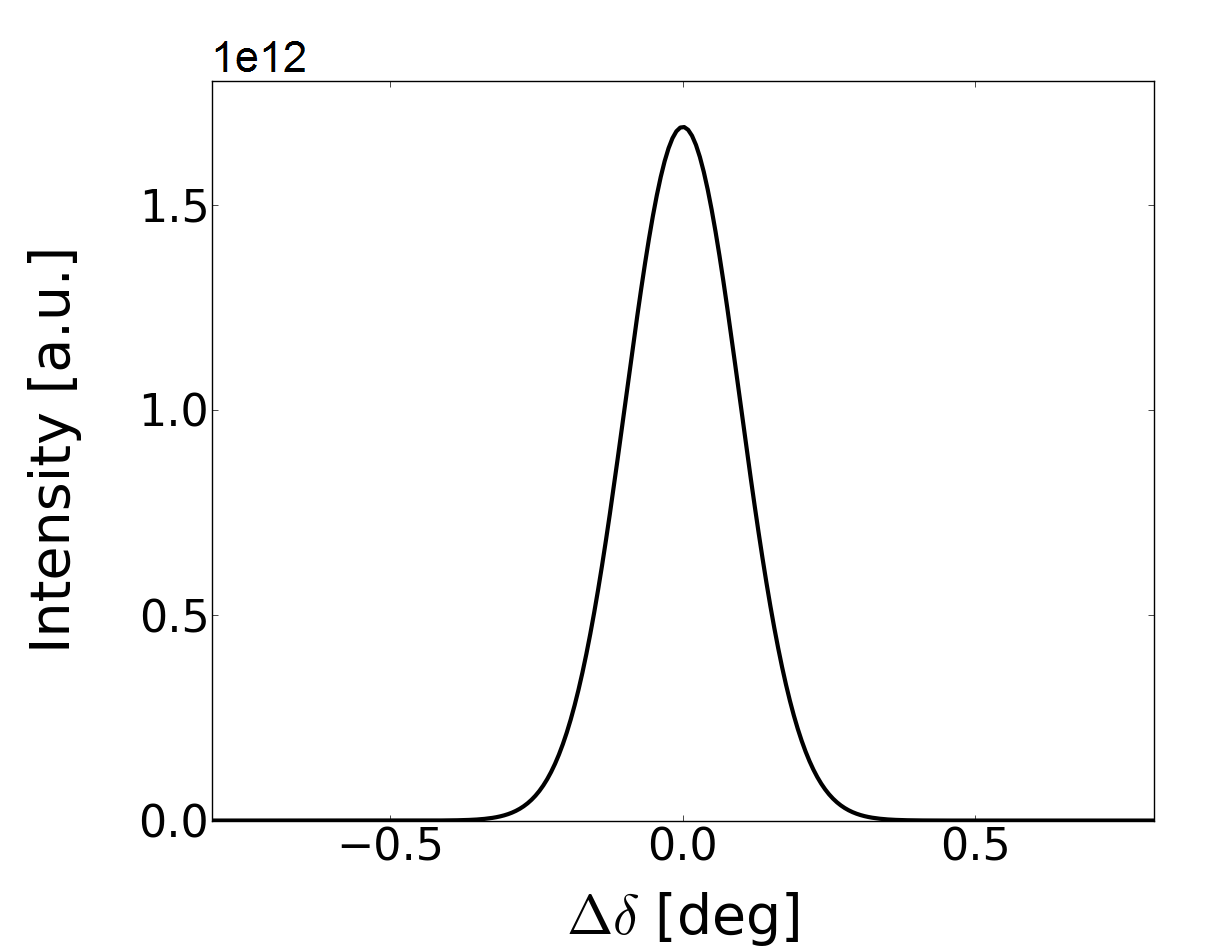}
	}
	\subfloat[Au(002)\label{Au_200__S010}]{
		\includegraphics[trim=0mm 5mm 0mm 0mm,width=0.2\textwidth]{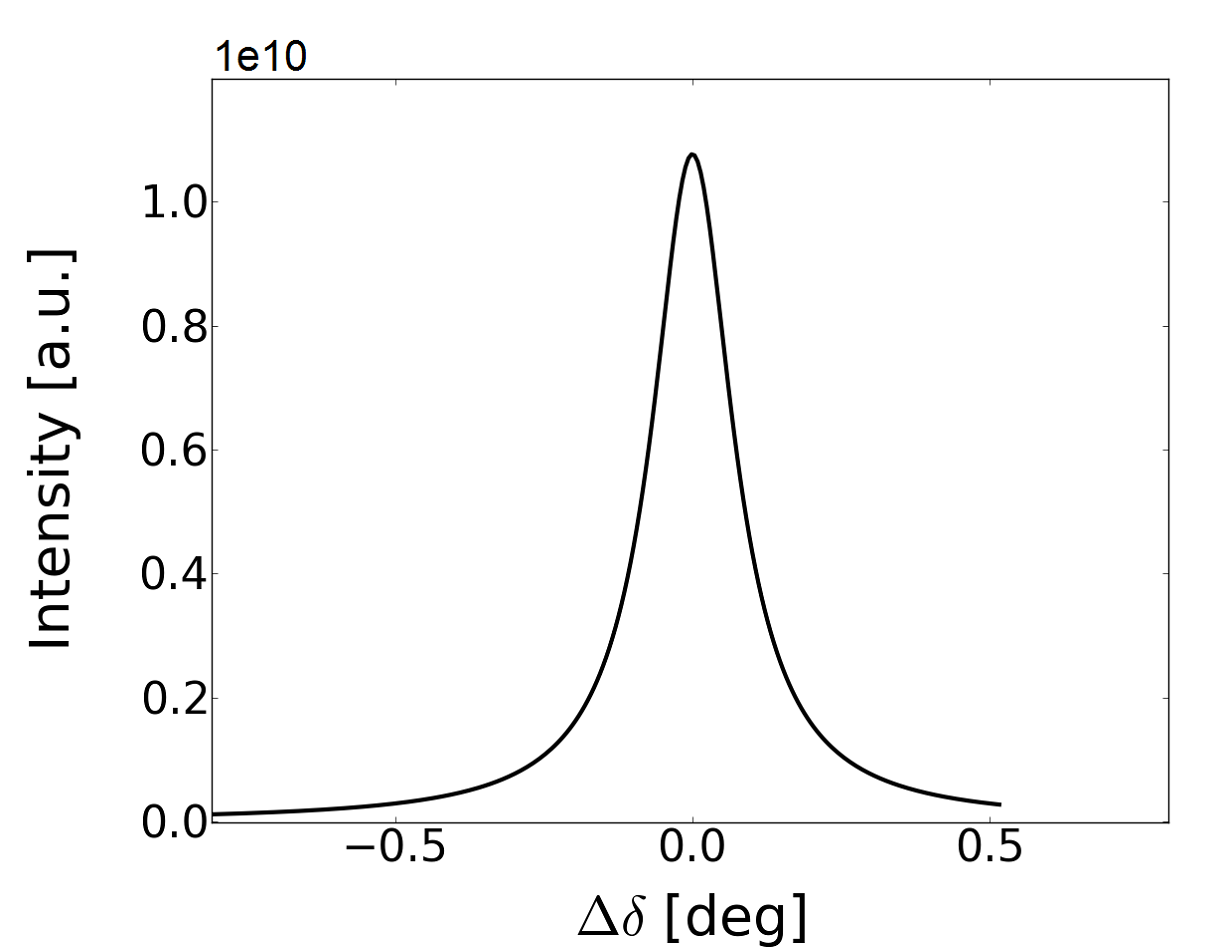}
	}
	\caption{Au peaks along the direction normal to the substrate surface, for the sample with 5 nm of Au layer thickness without STO thin film. The peaks were recorded through a detector scan ($\delta$ angle) of the diffractometer. Note the different intensity scale between the two peaks.}
	\label{normal_reflections_S010}
\end{figure}

Both the vertical orientations found for the Au nanoparticles were also clearly found in the sample without the STO layer, as can be seen in Fig.~\ref{normal_reflections_S010}. Therefore, as long as the direction normal to the substrate surface is considered, the crystallographic orientation of the Au nanoparticles seems to directly descend from that of the 3D islands forming the Au dewetted layer; in other words the deposition of the STO overlayer does not significantly modify the main Au crystallographic orientation (at least for an initial Au layer thickness around 5 nm).
Since the structure factor of all the allowed Au peaks is constant and equal to $4f$, $f$ being the atomic form factor, a rough indication of the abundance of the two orientations can be obtained evaluating the ratio of the corresponding diffracted intensities. An evaluation of the peaks areas led to a ratio $I_{111}/I_{002}\approx110$: the \{111\} orientation thus seems to be much more abundant than the \{200\} one. The same observation seems to be valid also for the samples with the STO thin film: the value of the ratio $I_{111}/I_{002}$ varies from one Au thickness value to another, but a large minimum value of 43.7 was in any case found. The presence of multiple Au growth directions is in agreement with the expected polycrystalline character of Au films grown over SrTiO$_3$(001) substrates and the high $I_{111}/I_{002}$ is consistent with the high Au surface energy anisotropy and diffusion coefficient on the substrate surface \cite{francis2007crystal}.

\subsection{In-plane orientation}\label{in_plane}
\subsubsection{Symmetry considerations}
\begin{figure*}
	\subfloat[\label{Symmetry_plot}]{
		\includegraphics[width=0.45\textwidth]{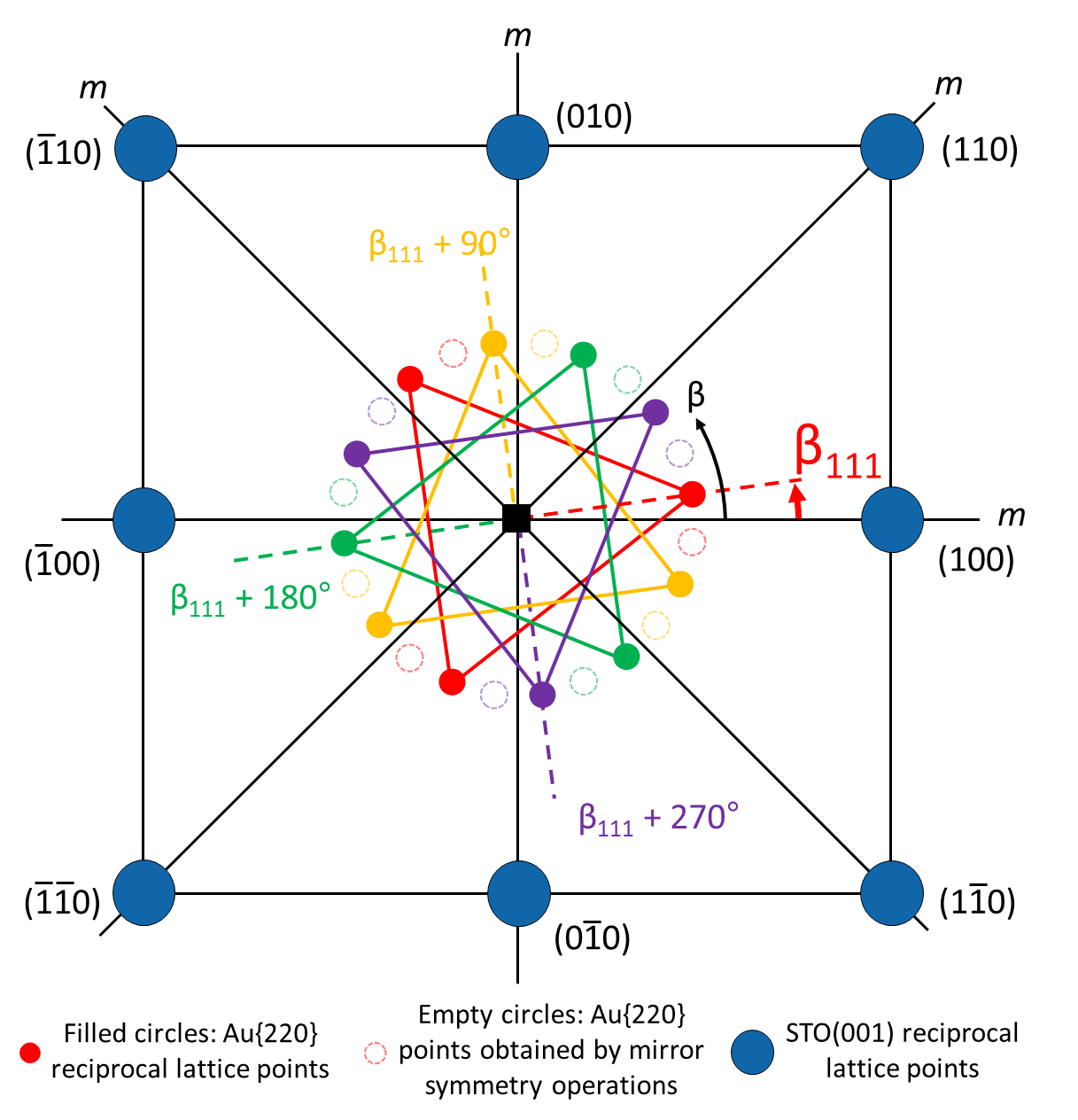}
	}
	\qquad
	\subfloat[\label{Symmetry_plot_002}]{
		\includegraphics[width=0.45\textwidth]{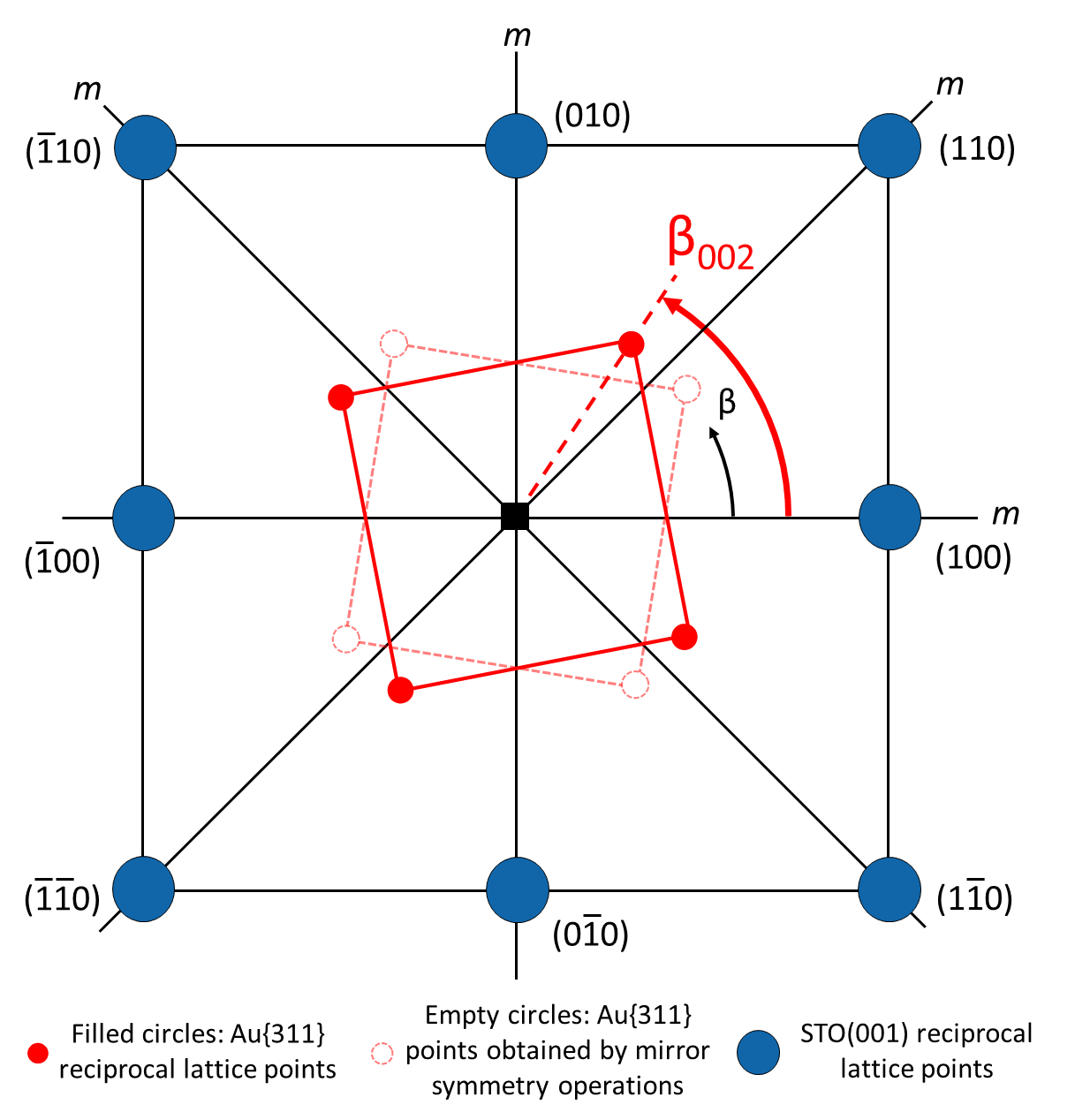}
	}
	\caption{(Color online) Possible orientations of the Au reciprocal lattice with respect to the STO reciprocal lattice for two different vertical (normal to the plane of the page) directions of the Au lattice. In both cases, the $\beta$ angle identifying the generic in-plane orientation of the reciprocal lattice points is reported in black. (a) Au\{220\} (or equivalently Au\{331\}) reciprocal lattice points as viewed along the Au(111): the in-plane orientation is defined through the angle $\beta_{111}$. (b) Au\{311\} reciprocal lattice points as viewed along the Au(002): the in-plane orientation is defined through the angle $\beta_{002}$.}
\end{figure*}
Once known the preferred nanoparticles orientation along the vertical (normal to substrate) direction, a complete characterization of the crystalline preferred orientations requires the investigation of the in-plane orientation, that is to say in a direction parallel to the substrate surface. For the experiment performed using the CCD area detector setup, this task was accomplished by measuring the periodicity of the Au\{220\} reflections upon a sample rotation around the vertical axis ($\phi$, or azimuthal, scan): only the in-plane orientation of the $(001)_{STO}\:||\:(111)_{Au}$ nanoparticles was studied, since they are much more abundant than the other family and thus more significant to investigate. For the diffractometer experiment both the nanoparticles family were probed. For the $(001)_{STO}\:||\:(111)_{Au}$ nanoparticles the periodicity of the Au\{331\} reflections (which have the same periodicity of the Au\{220\} around the (111) direction) was measured; for the $(001)_{STO}\:||\:(002)_{Au}$ nanoparticles the Au\{311\} reflections were exploited instead.

As viewed from the (111) direction, the Au\{220\} reciprocal lattice points describe an equilateral triangle and are related one to another by 120$^\circ$ rotations around the (111) direction. The in-plane orientation of the Au(111) lattice planes can be conveniently described introducing the $\beta_{111}$ angle: as shown in Fig.~\ref{Symmetry_plot}, it is defined as the angle formed by the substrate (100) direction and the direction passing through one of the Au\{220\} (or equivalently Au\{331\}) lattice points and the center of the equilateral triangle they form (which turns out to be the $(11\bar{2})$). Considering both the 3-fold symmetry of the Au\{220\} (or equivalently Au\{331\}) lattice points and the cubic symmetry of the STO lattice (with its mirror planes and the 4-fold rotation axis), the expected values for the $\beta$ in-plane orientation angle are given by the following combination
\begin{equation}
\label{eq:symmetry_combo}
\beta=\pm \beta_{111} + n\,90^\circ + m\,120^\circ \qquad \begin{array}{ll}n=0,1,2,3\\m=0,1,2\end{array}
\end{equation}
where $\pm$ accounts for the mirror symmetry of the STO(001) substrate, $n\,90^\circ$ for its 4-fold rotational symmetry and $m\,120^\circ$ for the 3-fold rotational symmetry of the Au(111) plane.
For the case of the $(001)_{STO}\:||\:(002)_{Au}$ nanoparticles, the Au\{311\} reciprocal lattice points describes a square, being related by $90^\circ$ rotations around the (002) axis. As for the previous case, the in-plane orientation can be defined by specifing the relative rotation of the Au crystal relative to the surrounding STO matrix: the corresponding angle is now called $\beta_{002}$. Considering again the cubic symmetry of the STO lattice the expected values for the $\beta$ in-plane orientation angle are given in this case by the following
\begin{equation}
\label{eq:symmetry_combo_002}
\beta=\pm \beta_{002} + n\,90^\circ \qquad n=0,1,2,3
\end{equation}
where $\pm$ accounts for the mirror symmetry of the STO(001) substrate and $n\,90^\circ$ for the 4-fold rotational symmetry of the Au(002) plane.

The equilibrium value of both the $\beta_{111}$ and the $\beta_{002}$ angle is expected to be mainly determined by energetic arguments. Both in the case of the Au 3D islands present prior to the STO thin film deposition and in the case of the embedded Au nanoparticles, the in-plane atomic arrangements will be the one which minimize the Au/STO interface energy. The situation is of course rather different depending on which of the two sytems, 3D islands or embedded nanoparticles, is considered. In the former case, the only interface present is the one between the (111) (or (002)) lattice planes of the islands and the single crystal substrate. In the latter case, on the other hand, the interface energy between the nanocrystal as a whole and the surrounding STO thin film has to be taken into account. The interface is thus more complicated, since it also depends on the crystallographic orientation of the terminating facets of each Au nanoparticles.

\subsubsection{Experimental results}
\begin{figure}
		\includegraphics[width=0.45\textwidth]{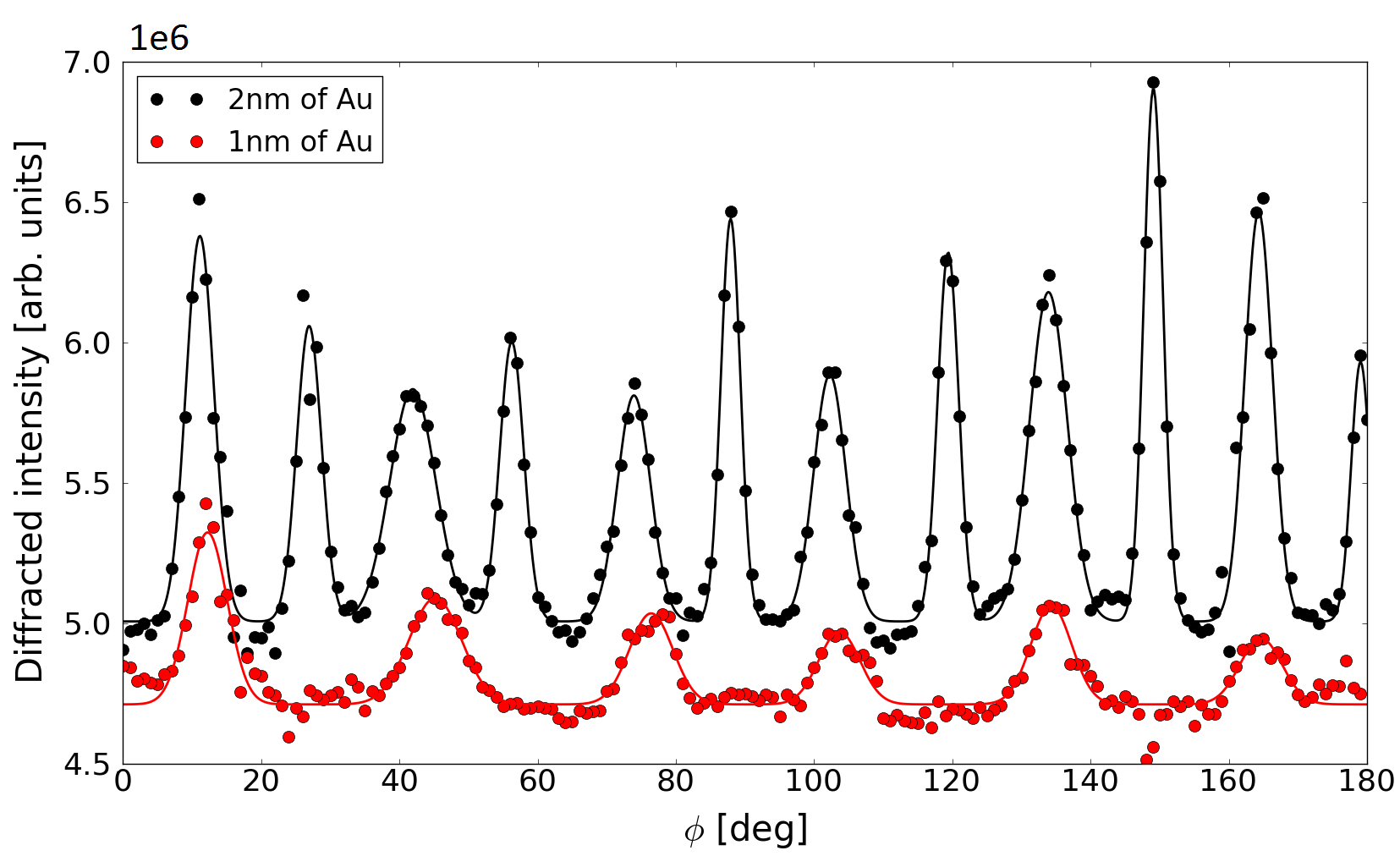}
		\caption{(Color online) Measured (circles) and gaussian-fitted (solid line) Au\{220\} diffracted intensity as a function of the $\phi$ azimuthal angle (directly related to the $\beta$ in-plane orientation angle except for a small offset in the sample mounting) for the samples with the two lowest initial Au layer thickness values. Lowering the Au seed layer thickness from 1 to 2 nm causes the peak periodicity to change.}
		\label{In_planeAu_111__Au_220__low_thickness}
\end{figure}

\begin{figure} 
		\includegraphics[width=0.45\textwidth]{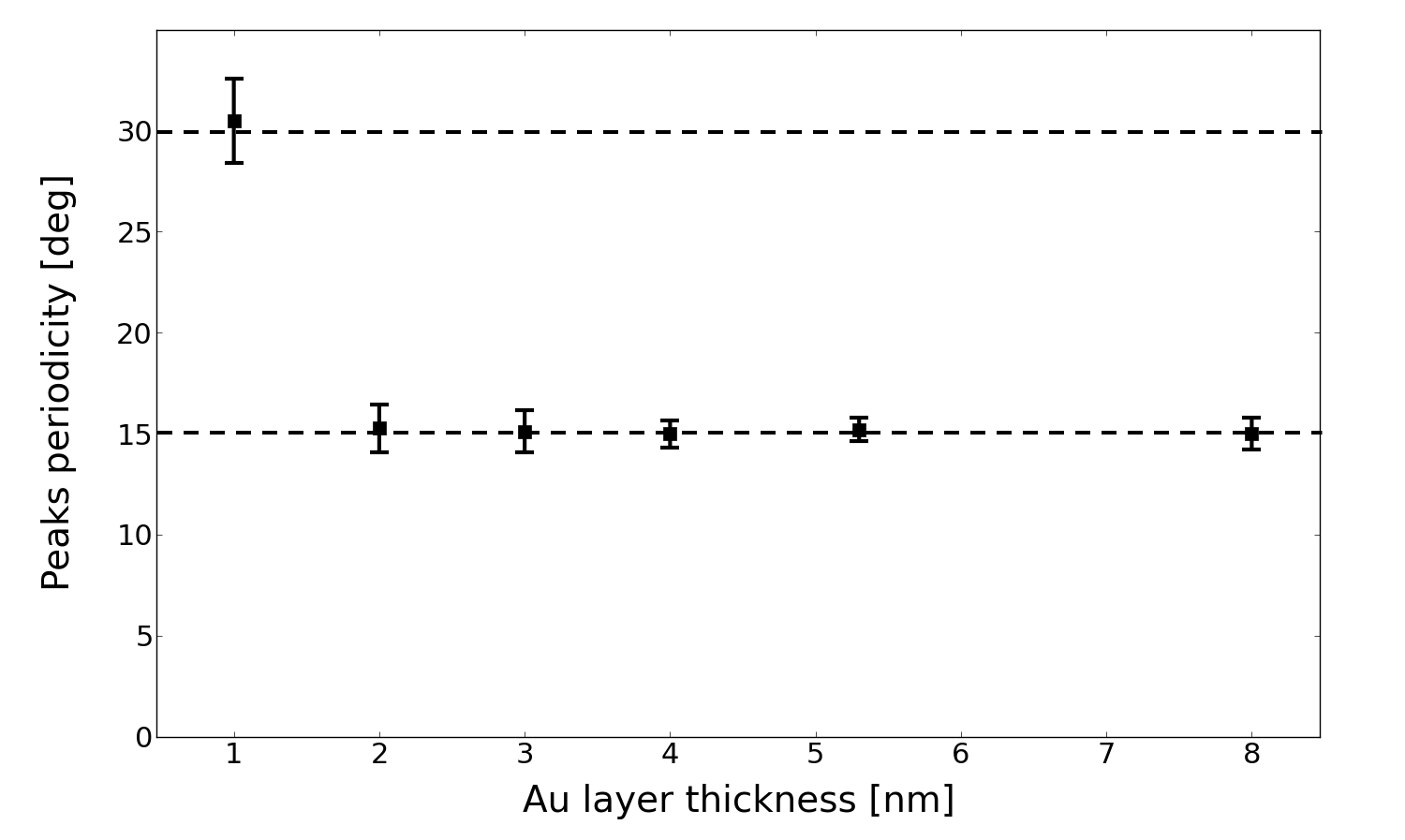}
		\caption{Periodicity of the Au\{220\} diffraction peaks over a scan of the $\phi$ angle as a function of the Au seed layer thickness: the periodicity abruptly changes from about $30^\circ$ to about $15^\circ$ between 1 and 2 nm of Au.}
		\label{Peaks_periodicity}
\end{figure}

\begin{figure}
		\includegraphics[width=0.45\textwidth]{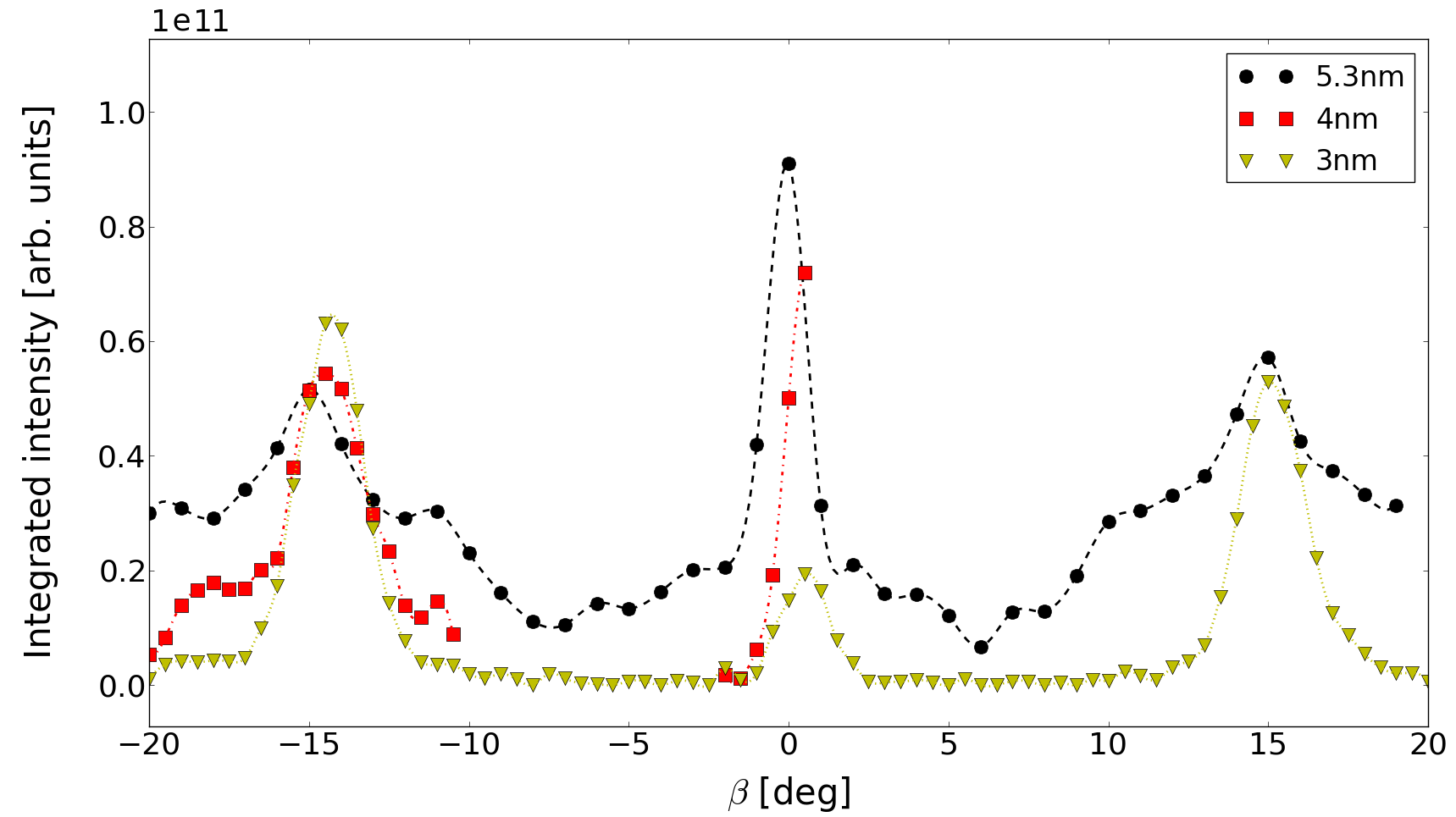}
		\caption{(Color online) Measured (symbols) and interpolated (dashed line) $(001)_{STO}\:||\:(111)_{Au}$ nanoparticles Au\{331\} diffracted intensity for different values of the initial Au layer thickness. The measured points refer to the diffracted intensity integrated over a rocking scan for the Au\{331\} reflections with different values of the $\beta$ in-plane orientation angle: the presence of the two nanoparticles in-plane orientation $\beta_{111}=0^\circ$ and $\beta_{111}=15^\circ$ is clearly shown. The dashed lines are just meant to help displaying the trend. The 4 nm data-set could not be completed ($\beta\leq1^\circ$). }
		\label{In_planeAu_111__Au_133__diffractometer}
\end{figure}

\begin{figure} 
\includegraphics[width=0.45\textwidth]{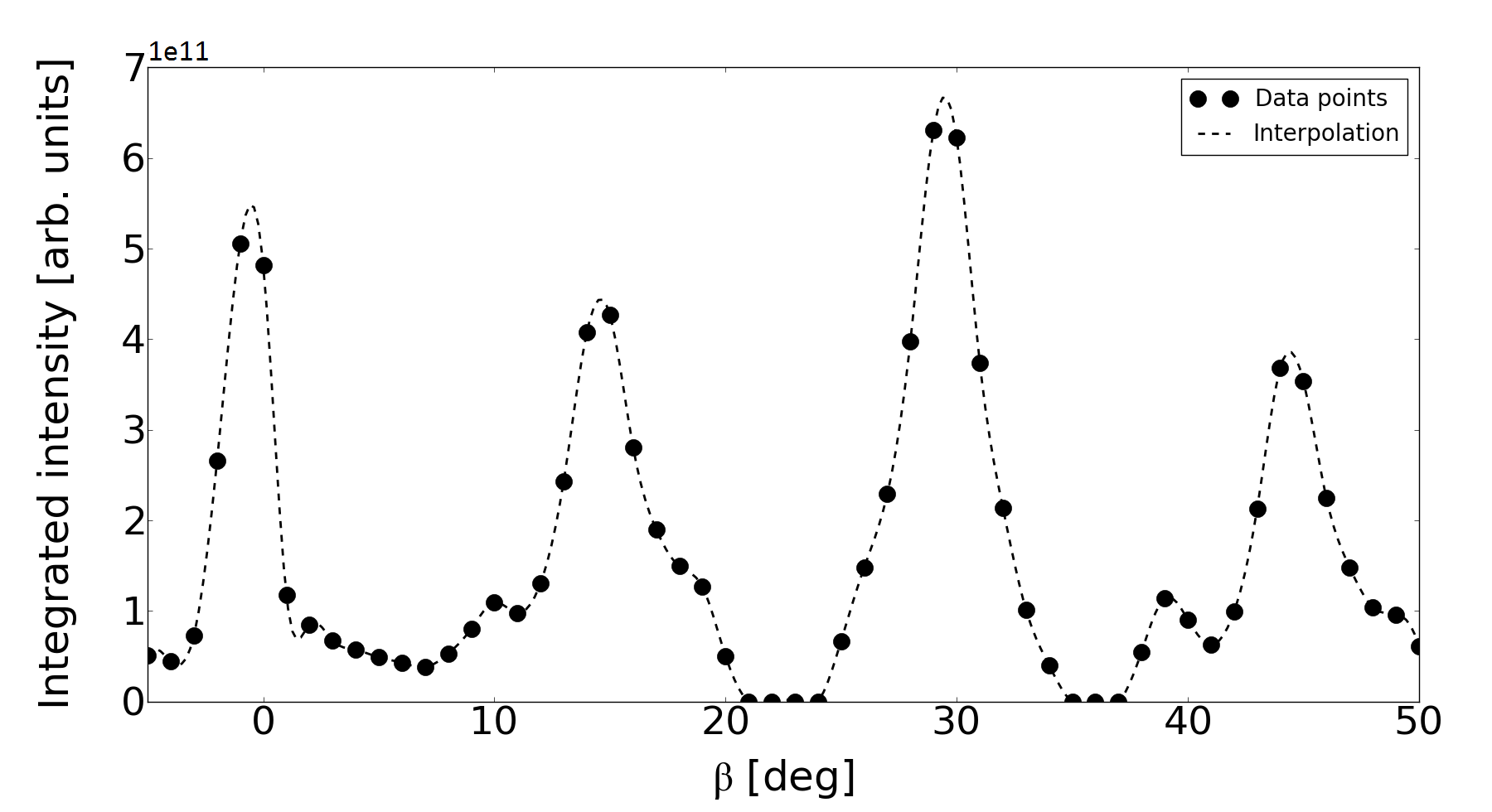}
\caption{Au\{331\} diffracted intensity of the $(001)_{STO}\:||\:(111)_{Au}$ 3D islands for the sample without STO thin film (5 nm of Au). The measured points refer to the diffracted intensity integrated over a rocking scan, as for the plots of Fig.~\ref{In_planeAu_111__Au_133__diffractometer}: the periodicity of the peaks is the same measured for the Au nanoparticles. The dashed line is just meant to help displaying the trend.}
\label{In_planeAu_111__Au_133__S010}
\end{figure}

Starting from the case of the $(001)_{STO}\:||\:(111)_{Au}$ nanoparticles, the Au\{220\} peaks show roughly the same periodicity varying the initial Au layer thickness from 8 nm down to 2 nm. An abrupt change seems to occur for the sample with 1 nm of Au, whose corresponding plot is shown in Fig.~\ref{In_planeAu_111__Au_220__low_thickness} together with the 2 nm one for comparison: from the figure is clearly evident how the number of peaks for 1 nm of Au is halved with respect to the 2 nm case, one peak every two being missing.
Fitting the data points with a train of gaussian curves, a practically identical peaks periodicity value of about $15^\circ$ was obtained for all the sample between 8 and 2 nm of Au. On the contrary (see Fig.~\ref{Peaks_periodicity}), the sample with 1 nm of Au diplays a double periodicity, with a peak-to-peak distance of about $30^\circ$. An evaluation of expression~(\ref{eq:symmetry_combo}) leads to the conclusion that the presence of equally spaced peaks with a periodicity of $15^\circ$ in the azimuthal scans can be accounted by the presence of two families of Au nanoparticles, one with $\beta_{111}=0^\circ$ and the other with $\beta_{111}=45^\circ$ (or, equivalently, $\beta_{111}=15^\circ$), being $\beta_{111}$ the angle defined in Fig.~\ref{Symmetry_plot}. Although the absolute $\phi$ values at which the peaks of Fig.~\ref{In_planeAu_111__Au_220__low_thickness} occur are directly related to the $\beta$ in-plane angle of Fig.~\ref{Symmetry_plot}, they are affected by a small unknown offset in the sample mounting. 
This uncertainty was removed through the diffractometer data shown in Fig.~\ref{In_planeAu_111__Au_133__diffractometer}, which clearly shows the existence of the two nanoparticles in-plane orientations $\beta_{111}=0^\circ$ and $\beta_{111}=15^\circ$.
The above conclusion would immediately lead to interpret the double periodicity observed for the 1 nm sample as a direct result of the disappearance of one of the two different nanoparticles in-plane orientations. Even considering the small unknown offset in the $\phi$ angle values, the family which seems to be absent in the sample with 1 nm of Au (Fig.~\ref{In_planeAu_111__Au_220__low_thickness}) is the one characterized by $\beta_{111}=0^\circ$. The peaks sequence observed is indeed very close to $15^\circ,45^\circ,75^\circ,\ldots$, produced by $\beta_{111}=15^\circ$. This seems to be elegantly predicted by the variation in the peaks intensity of Fig.~\ref{In_planeAu_111__Au_133__diffractometer}. As can be clearly seen, the intensity of the central peak drops as the Au thickness decreases: while it is the dominant peak for 5.3 nm of Au, in the 3 nm sample the intensity associated with the $\pm15^\circ$ peaks becomes much greater. This trend leads to infer that further lowering of the Au layer thickness would lead to the total suppression of the central peak in favor of the other two. The same nanoparticles in-plane orientation for the 8-2 nm thickness range was also found in the sample without STO top layer (Fig.~\ref{In_planeAu_111__Au_133__S010}). This is consistent with the epitaxial orientations of Au islands on a SrTiO$_3$(001)~-~(2~x~1) surface found by \citet{silly2006bimodal}: in particular, the cases $\beta_{111}=0^\circ$ and $\beta_{111}=45^\circ$ correspond to the $(001)_{SrTiO_3} \: || \: (111)_{Au}\;,\; [100]_{SrTiO_3} \: || \: [110]_{Au}$ and $(001)_{SrTiO_3} \: || \: (111)_{Au} \;,\; [110]_{SrTiO_3} \: || \: [110]_{Au}$ orientations they report, respectively. As for the preferred nanoparticles vertical growth orientation, the nanoparticles in-plane orientation thus seems to directly derive from that of the Au layer 3D islands. The available data does not allow to definitively establish whether the intensity drop of the $\beta_{111}=0^\circ$ peak and the disappearance of the corresponding nanoparticles family is induced by the growth of the STO layer or is already present in the Au 3D islands. However, considering the lack of documented crystallographic orientations  thickness dependences in traditional layered samples, a role of the overgrown thin film is likely to be supposed.

\begin{figure} 
\includegraphics[width=0.45\textwidth]{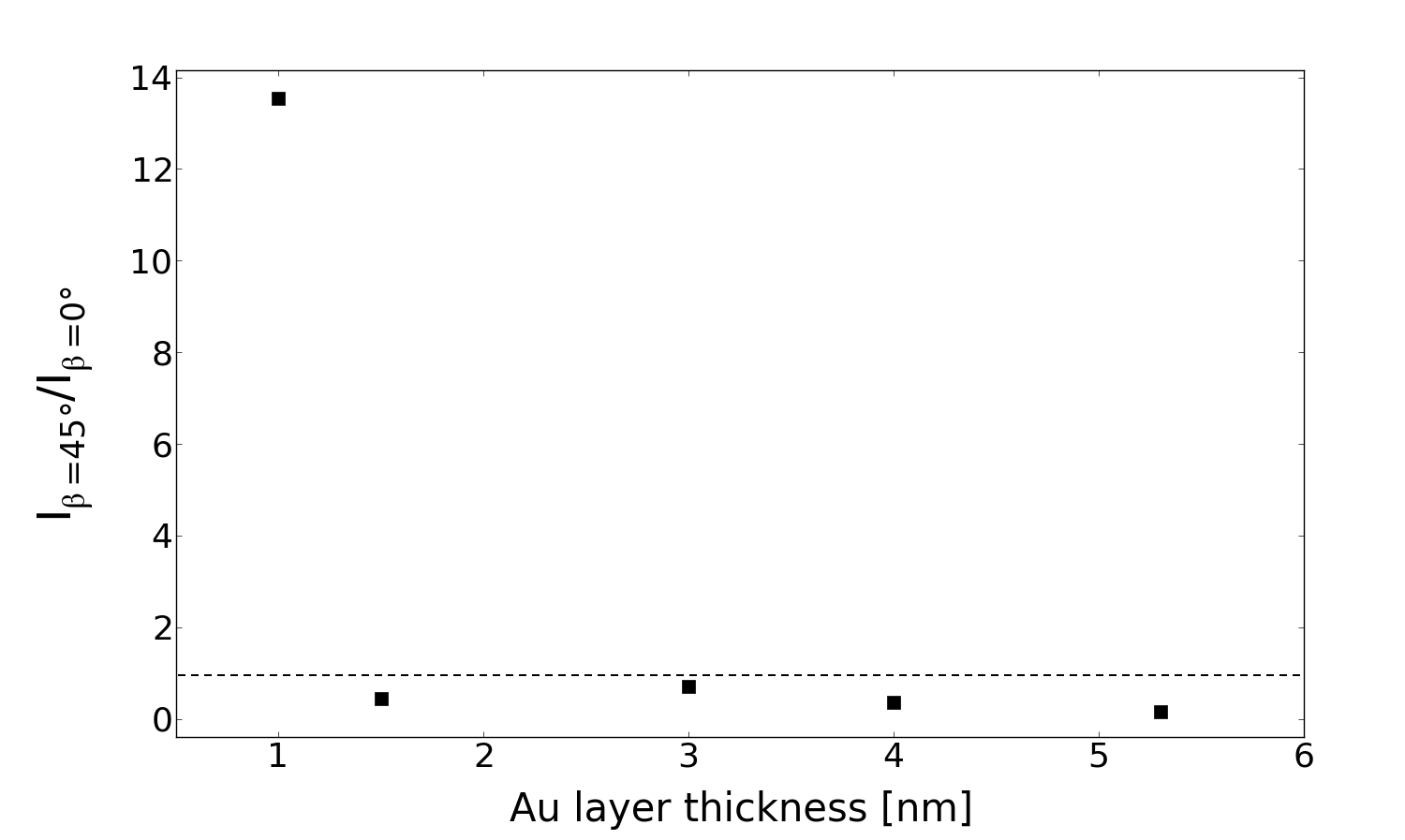}
\caption{Diffracted intensity ratio between the Au\{311\} peaks at $\beta=45^\circ$ and the same peaks at $\beta=0^\circ$ for different initial Au layer thicknesses. The ratio provides an indication about the relative abundance of the $\beta_{002}=45^\circ$ family with respect to the $\beta_{002}=0^\circ$ one: an abrupt change in the ratio value occur for the 1 nm sample. The values used to calculate the ratio are the maximum intensity values obtained from the plots of Fig.~\ref{002_contour_plots}.}
\label{002_ratio}
\end{figure}

\begin{figure} 
	\includegraphics[width=0.4\textwidth]{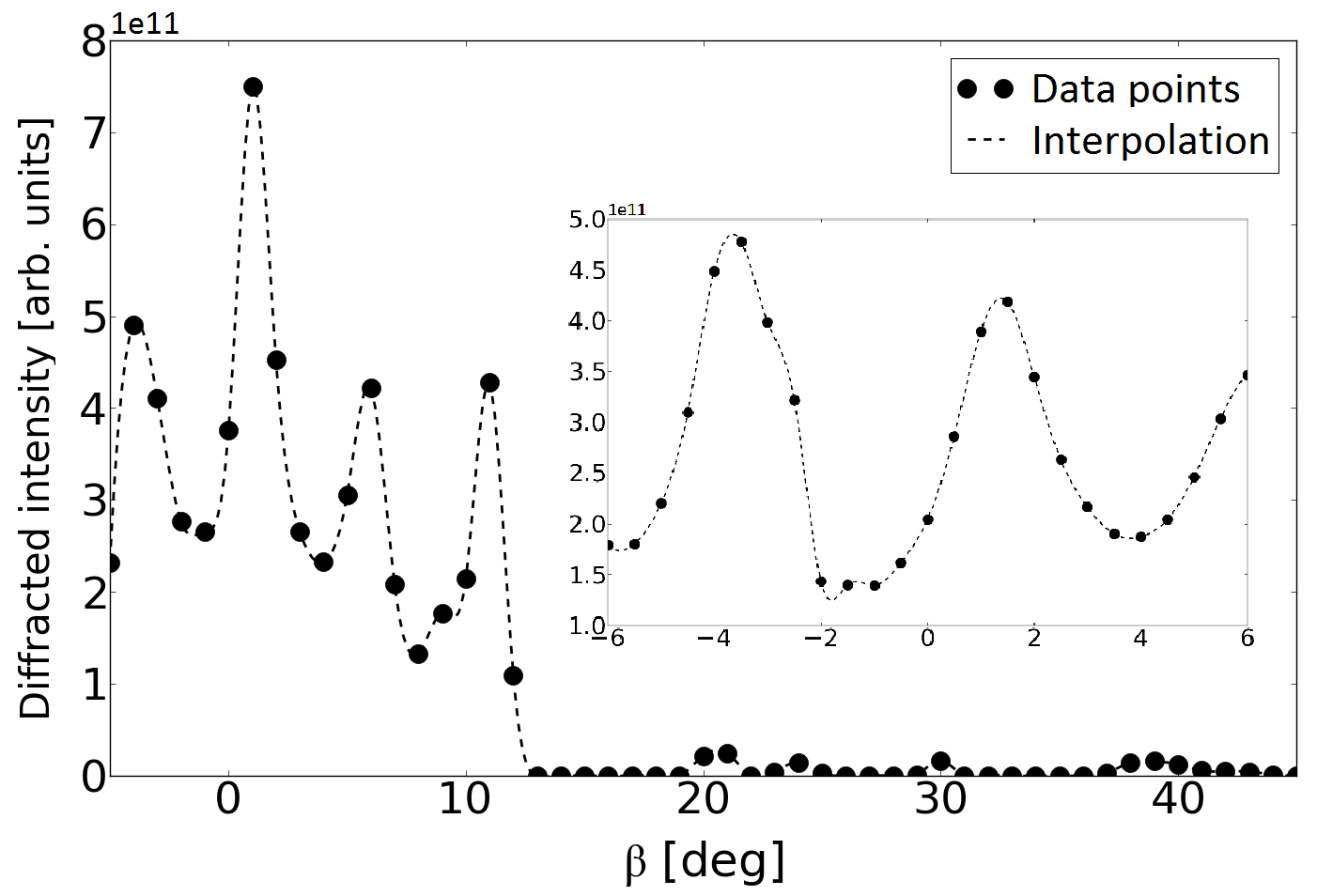}
	\caption{Au\{311\} diffracted intensity of the $(001)_{STO}\:||\:(002)_{Au}$ 3D islands for the sample without STO thin film (5 nm of Au). The measured points refer to the diffracted intensity integrated over a rocking scan. Multiple peaks are present in the interval $(-6^\circ,10^\circ)$; in the inset, a finer scan around $\beta=0^\circ$ is shown. The dashed line is just meant to help displaying the trend.}
	\label{In_planeAu_002__Au_113__S010_total}
\end{figure}

\begin{figure*}
	\subfloat[5.3 nm\label{Contourf_In_planeAu_002__Au_113__HB18A}]{
		\includegraphics[trim=0mm 10mm 0mm -10mm,width=0.5\textwidth]{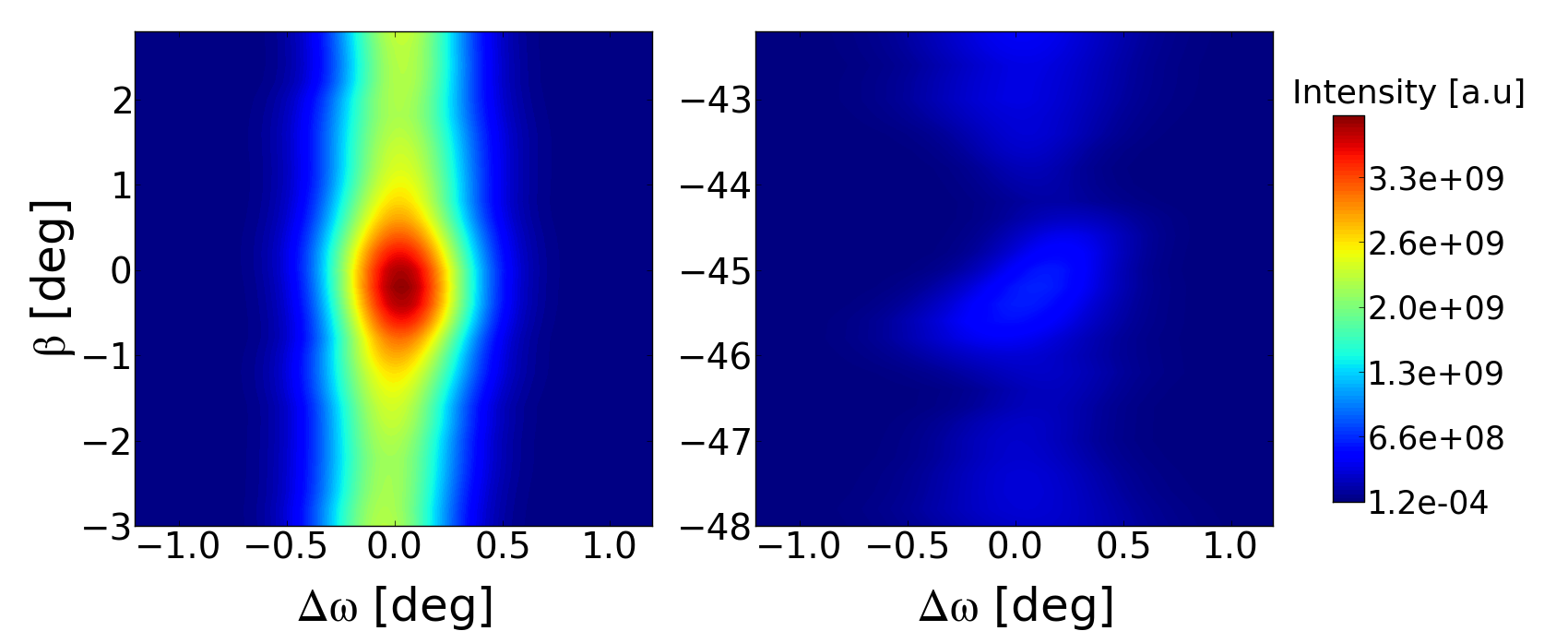}
	}
	\subfloat[4 nm\label{Contourf_In_planeAu_002__Au_113__HB17A}]{
		\includegraphics[trim=0mm 10mm 0mm -10mm,width=0.5\textwidth]{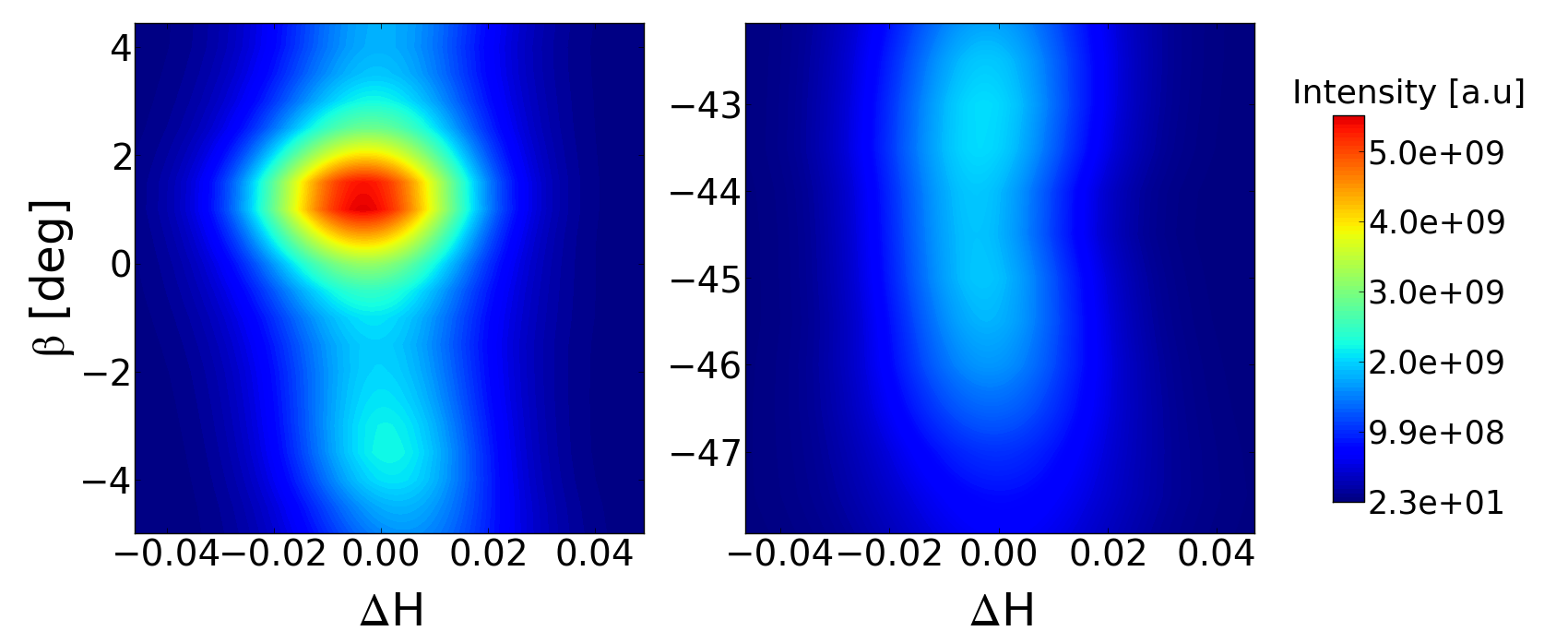}
	}
	\hfill
	\subfloat[3 nm\label{Contourf_In_planeAu_002__Au_113__HB16A}]{
		\includegraphics[trim=0mm 10mm 0mm -10mm,width=0.5\textwidth]{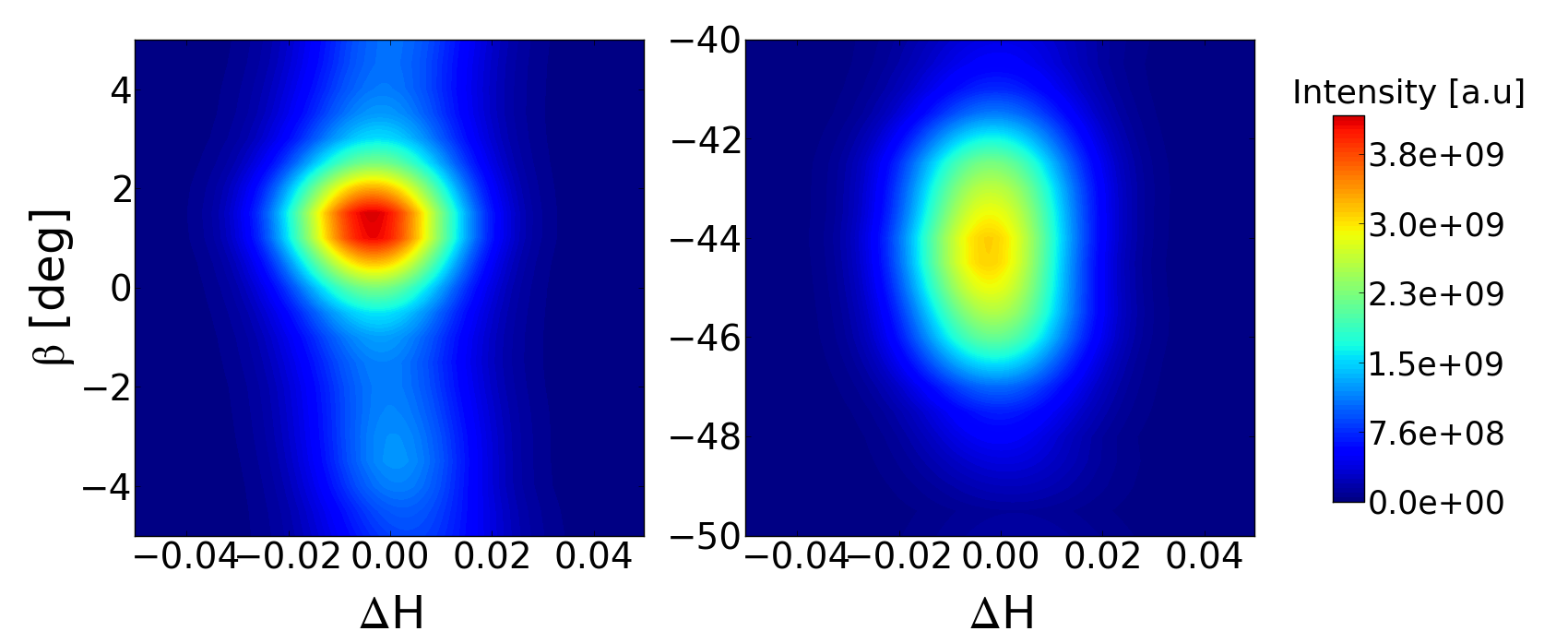}
	}
	\subfloat[1.5 nm\label{Contourf_In_planeAu_002__Au_113__S043}]{
		\includegraphics[trim=0mm 10mm 0mm -10mm,width=0.5\textwidth]{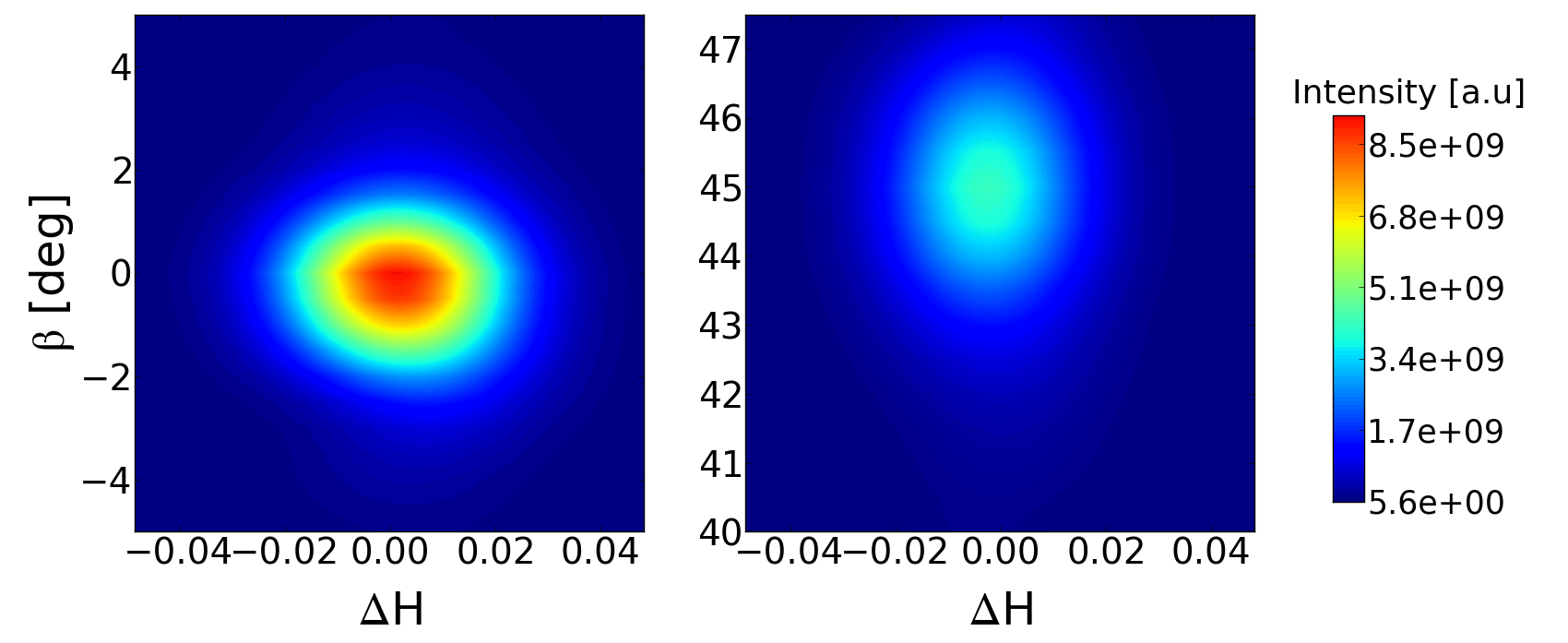}
	}
	\hfill
	\subfloat[1 nm\label{Contourf_In_planeAu_002__Au_113__HB14A}]{
		\includegraphics[trim=0mm 10mm 0mm -10mm,width=0.5\textwidth]{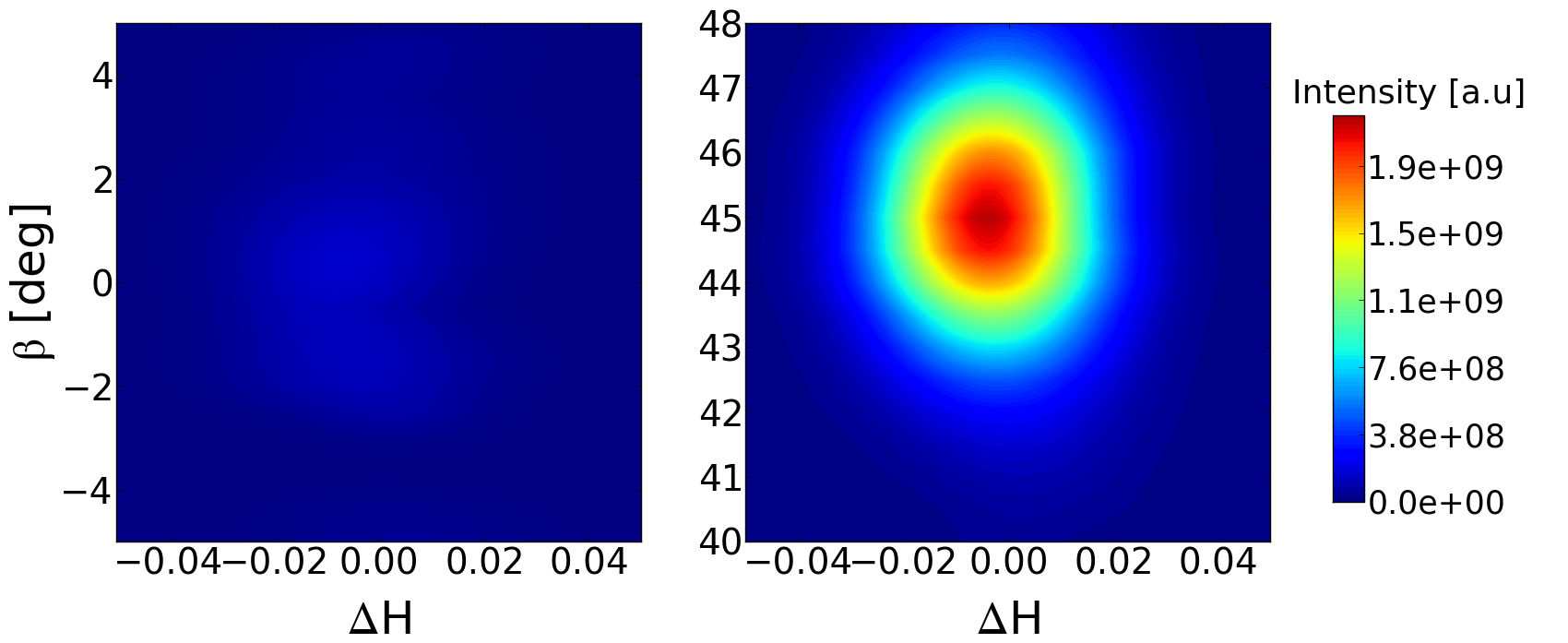}
	}
	\caption{\label{002_contour_plots}(Color online) Au\{311\} rocking scans for different $(001)_{STO}\:||\:(002)_{Au}$ nanoparticles in-plane orientations ($\beta$ values) around $\beta=0^\circ$ (left) and $\beta=45^\circ$ (right). The data refers to samples with different values of the Au seed layer thickness: the left and the right plot of each thickness share the same colorscale to allow a direct comparison. In the x-axis is reported either the variation of the $\omega$ sample rocking angle or the corresponding variation of one of the Miller indices in the plane of the substrate surface. The change in the intensity distribution between the left and the right peak occurring for the 1 nm sample is shown.}
\end{figure*}

With respect to the $(001)_{STO}\:||\:(002)_{Au}$ nanoparticles, the data collected with the diffractometer (see Fig.~\ref{002_contour_plots}) show the presence of Au\{311\} peaks at values of the $\beta$ in-plane orientation angle equal to $0^\circ$ and $45^\circ$. This leads to the conclusion that, even in this case, two nanoparticles families with a different in-plane orientation are present: these two in-plane orientations are characterized by $\beta_{002}=0^\circ$ and $\beta_{002}=45^\circ$ respectively (being $\beta_{002}$ the angle defined in Fig.~\ref{Symmetry_plot_002}). They correspond to the epitaxial relations $(001)_{SrTiO_3} \: || \: (001)_{Au}\;,\; [100]_{SrTiO_3} \: || \: [110]_{Au}$ and $(001)_{SrTiO_3} \: || \: (001)_{Au} \;,\; [100]_{SrTiO_3} \: || \: [100]_{Au}$ respectively, documented in literature for a generic FCC or BCC metal deposited on a SrTiO$_3$(001) substrate \cite{fu2007interaction}. 
However, as for the $(001)_{STO}\:||\:(111)_{Au}$ nanoparticles in-plane orientation, a transition seems to occur by lowering the Au seed layer thickness toward the 1 nm value. While for the range 1.5-5.3 nm the $\beta=0^\circ$ Au\{311\} peak is the most intense one, the 1 nm sample displays an opposite trend, the $\beta=0^\circ$ peak being almost absent. This result is summarized in the plot of Fig.~\ref{002_ratio}, where the intensity ratio $I_{\beta=45^\circ}/I_{\beta=0^\circ}$ is shown as a function of the Au layer thickness.
With respect to the $(001)_{STO}\:||\:(111)_{Au}$ nanoparticles, the relation between the orientation of the Au 3D islands and the nanoparticles one is in this case less obvious. In the case of the 3D islands multiple peaks are indeed present in the $(-6^\circ,10^\circ)$ interval (Fig.~\ref{In_planeAu_002__Au_113__S010_total}). The STO thin film thus seems somehow to force only one of the in-plane orientations present prior to its deposition to occur for the embedded nanoparticles. It's interesting how none of these peaks is exactly centered at $\beta=0^\circ$: this evidence might be related to the small deviations from the perfect $\beta=0^\circ$ position observed for the nanoparticles peaks (left plots of Fig.~\ref{002_contour_plots}). Furthermore, contrary to what measured for the Au nanoparticles, no peak seems to be present around $\beta=45^\circ$ for the 3D islands. However, the absence of this peak could be considered consistent with the relatively low intensity (the smallest measured) of the $45^\circ$ peak for the sample with the STO layer and a comparable Au seed layer thickness (5.3 nm, compared to the 5 nm of the sample without STO layer). Once again, the role of the STO layer in the appearance of the differences mentioned above, as well as in the occurence of the transition for the 1 nm sample, has to be further investigated. 

\clearpage

\section{Conclusions and future developments}
Au nanoparticles are promising as plasmonic active materials, with potential applications as sensors in life sciences. Samples were prepared by depositing a Au layer on a SrTiO$_3$(001) single crystal surface, annealing the Au layer (causing it to dewet into Au islands) and capping the dewetted layer with a thin film of SrTiO$_3$. During the last deposition step, the Au dewetted layer self-assembles to homogeneously arranged monocrystalline nanoparticles mostly located at a characteristic depth within the deposited SrTiO$_3$ thin film. Synchrotron x-ray diffraction was used to investigate the epitaxial relation of the embedded Au nanoparticles with the SrTiO$_3$ matrix. 

The collected data revealed the presence of two main orientations of the nanoparticles along the STO(001) surface normal, namely (111) and (002), with the (111) being predominant. For these out-of-plane orientations the in-plane orientation was studied.  For the dominant nanoparticles with $(001)_{SrTiO_3} \: || \: (111)_{Au}$, two in-plane orientations with  $[110]_{SrTiO_3} \: || \: [110]_{Au}$ and $[100]_{SrTiO_3} \: || \: [110]_{Au}$ were found in the samples with an initial Au layer thickness of 2 nm and above, as observed previously in similar, non-capped systems  \cite{silly2006bimodal}. For the nanoparticles with $(001)_{SrTiO_3} \: || \: (002)_{Au}$, the two orientations $[100]_{SrTiO_3} \: || \: [100]_{Au}$ and $[100]_{SrTiO_3} \: || \: [110]_{Au}$ were found, again in good agreement with the literature \cite{fu2007interaction}. In general, the orientation exhibited by the Au nanoparticles seem to be strongly influenced by the one measured for the Au 3D islands present on the substrate surface prior to the STO thin film growth. 

For the first time, strong evidence of a transition in the nanoparticles crystalline preferred orientation occurring  between 1 and 2 nm of Au layer thickness was found. For the 1 nm sample, the $(001)_{SrTiO_3} \: || \: (111)_{Au}\;,\; [100]_{SrTiO_3} \: || \: [110]_{Au}$ orientation apparently disappear and the relative abundance of the $(001)_{SrTiO_3} \: || \: (002)_{Au}\;,\; [100]_{SrTiO_3} \: || \: [110]_{Au}$ orientation decreases with respect to the $(001)_{SrTiO_3} \: || \: (002)_{Au} \;,\; [100]_{SrTiO_3} \: || \: [100]_{Au}$ one. A previously unreported dependence of the Au nanoparticle’s preferred orientations on the layer thickness has thus been unveiled.

\begin{acknowledgments}
We would like to thank J.P. Wright and C. Ferrero for the valuable support in the analysis of the 2D diffraction data and E. Chinchio for having provided the detector and for his support in the data treatment. We also thank R. Felici for critical reading of the manuscript. A special thank you to H. Simons for the support provided during all the work and for the help in the draft of this paper.
\end{acknowledgments}

\bibliography{References}

\end{document}